\documentclass[a4paper,fleqn,usenatbib]{mnras}

\usepackage[T1]{fontenc}
\usepackage{ae,aecompl}

\usepackage{graphicx}	
\usepackage{amsmath}	
\usepackage{amssymb}	

\title[Dark matter cores in EAGLE]{Baryon-induced dark matter cores in the EAGLE simulations}

\author[Alejandro Ben\'itez-Llambay]{
Alejandro Ben\'itez-Llambay,$^{1}$\thanks{E-mail: alejandro.b.llambay@durham.ac.uk (ABL)}
Carlos S. Frenk,$^{1}$
Aaron D. Ludlow$^{2}$ \&
\newauthor{Julio F. Navarro$^{3}$}
\\
$^{1}$Institute for Computational Cosmology, Department of Physics, Durham University, South Road, Durham, DH1 3LE, UK\\
$^{2}$International Centre for Radio Astronomy Research, University of Western Australia, 35 Stirling Highway,
Crawley, \\ Western Australia 6009, Australia\\
$^{3}$Senior CIfAR Fellow. Department of Physics \& Astronomy, University of Victoria, BC, V8P 5C2, Canada}

\date{Accepted XXX. Received YYY; in original form ZZZ}

\pubyear{2015}

\begin{document}
\label{firstpage}
\pagerange{\pageref{firstpage}--\pageref{lastpage}}
\maketitle

\begin{abstract}
  We examine the formation of dark matter (DM) cores in dwarf galaxies
  simulated with the EAGLE model of galaxy formation. As in earlier
  work, we find that the star formation (SF) gas density threshold ($\rho_{\rm th}$) plays a critical role. At low thresholds (LT), gas is unable to reach densities high enough to dominate the gravitational potential before being dispersed by feedback from supernovae. LT runs show little effect on the inner DM profile, even in systems with extended and bursty SF, two ingredients often cited as critical for core
  formation. For higher thresholds, gas is able to dominate the
  gravitational potential before being ejected by feedback. This can
  lead to a substantial reduction in the inner DM content, but only if
  the gas is gravitationally important over an extended period of
  time, allowing the halo to contract before gas removal. Rapid assembly
  and removal of gas in short SF bursts is less effective at
  altering the inner DM content. Subsequent gas accretion may draw DM
  back in and reform a cusp, unless SF is bursty enough to prevent it,
  preserving the core. Thus, for the EAGLE SF+feedback model, there is no
  simple relation between core formation and SF history, contrary to
  recent claims.
  The dependence of the inner DM content of dwarfs on $\rho_{\rm th}$
  hinders robust predictions and the interpretation of observations. A simulation of a $(12 \rm \ Mpc)^3$ volume with high $\rho_{\rm th}$ results in dwarfs with sizeable cores over a limited halo mass range, but with insufficient variety in mass profiles to explain the observed diversity of dwarf galaxy rotation curves.
\end{abstract}

\begin{keywords}
galaxies: dwarf -- galaxies: haloes -- (cosmology: dark matter)
\end{keywords}



\section{Introduction}
\label{Sec:Introduction}

A fundamental prediction of cosmological collisionless N-body
simulations of structure formation is that matter assembles into
gravitationally bound haloes whose density profile approximately
follows the universal Navarro-Frenk-White (hereafter, NFW) form,
independently of initial conditions and cosmological
parameters~\citep{Navarro1996a,Navarro1997}.  This density law
diverges towards the centre as $r^{-1}$ and falls off in
the outer parts as $r^{-3}$.

In contrast to this well established result, measurements of rotation
curves and dynamical modelling of nearby low-mass galaxies have often
been claimed to require much shallower (or even flat) inner density
profiles~\citep[e.g.,][and references therein]{Moore1994, Flores1994,
  Walker2011, Oh2015}, although these claims have been recently
disputed~\citep{Strigari2014, Pineda2017, Oman2017, Genina2018}. The
apparent discrepancy between collisionless simulations and
observations is the ``core-cusp'' problem.

From a theoretical point of view, the existence of shallower than NFW
density profiles in low-mass galaxies has sometimes been interpreted as the
manifestation of dark matter self-interactions~\cite[e.g.,][and
references therein]{Spergel2000, Bullock2017}. However, generating
shallower density profiles is not particularly challenging for the
$\Lambda$ Cold Dark Matter ($\Lambda$CDM) cosmological model. Indeed,
it was already recognized over twenty years ago that baryonic
processes could, in principle, transform a central dark matter cusp
($\rho \propto r^{-1}$) into a core ($\rho \propto r^{0}$;
\citealt{Navarro1996b})\footnote{This does not necessarily imply that
  simulated cores are comparable to those inferred by
  observations.}. Various specific mechanisms have been proposed to
achieve this~\citep[see, e.g.,][for an overview]{deBlok2010}, but sudden loses of baryonic material from central
regions of haloes is perhaps the most widely accepted. Using N-body
simulations,~\cite{Navarro1996b} showed that the sudden removal of a
large amount of baryons from the centre of a cuspy dark matter halo
leads to the formation of a core. This idea was later confirmed and
extended using a variety of codes and numerical
setups~\citep[e.g.,][]{Gelato1999, Read2005, Governato2010,
  Pontzen2012, Teyssier2013}.

Particularly relevant is the work of~\cite{Pontzen2012} who used
cosmological zoom-in simulations of the formation of dwarf galaxies to
show that not only very violent~\citep[as suggested
by][]{Navarro1996b}, but also moderate and repeated perturbations to
the gravitational potential in the inner regions of dark matter haloes
can significantly shallow the central cusp. This result confirmed
that large-scale supernovae-driven winds may significantly affect the
inner structure of dark matter haloes.

Subsequent cosmological simulations performed with different codes
such as GASOLINE~\citep[e.g.,][]{Governato2010, Zolotov2012,
  DiCintio2014a, Tollet2016} or FIRE~\citep[e.g.][]{Chan2015,
  Wetzel2016, Fitts2017, Hopkins2017} have reinforced the idea that
baryonic blowouts can perturb the inner regions of dark matter
haloes. These simulations have shown that cores form more efficiently
in a relatively narrow range of stellar/halo mass that matches roughly
that of nearby bright dwarf galaxies in the Local Group
(LG). The simulations also suggest that the inner slope of dark matter
haloes may be mass dependent~\citep{Governato2012, DiCintio2014b}. 

In sharp contrast, simulations performed with other codes such as
  EAGLE~\citep[e.g.][]{Schaye2015, Oman2017, Genina2018} or
  AREPO~\citep{Vogelsberger2014} have shown that galaxies,
irrespective of their stellar or halo mass, always reside in cuspy NFW
haloes. This is certainly controversial: simulations
from~\cite{Zolotov2012} or the Latte project~\citep{Wetzel2016}
report central cores in dwarf galaxies, whereas the
  APOSTLE~\citep{Sawala2016} and AURIGA~\citep{Grand2016}
projects do not, yet they all claim a reasonably good agreement
between structural and statistical properties of the simulated
population of dwarfs and nearby LG dwarfs~\citep{Sawala2016,
  Simpson2017}.

Why then do simulated dwarfs with similar overall properties differ so
much in their inner mass distribution? One possibility are numerical
errors; another is the subgrid physics
modelling. The latter has been extensively invoked in the literature
to explain the discrepancies among simulations, although with little
quantitative support. For example, different degrees of ``burstiness''
in the star formation histories of simulated dwarfs is often argued
as the source of the discrepancy ~\citep[see e.g.,][and references
therein]{Chan2015, Onorbe2015, Bullock2017}. However, the efficient
stellar feedback implemented in the EAGLE and AURIGA codes
produces star formation histories with a similar degree of
``burstiness''~\citep[e.g.][]{Sharma2016} as those reported by,
e.g.,~\cite{Fitts2017}. Yet {\it none} of the EAGLE haloes
develop ``cores''~\citep{Bose2018}.

A clue to understanding the differences amongst simulations was
already pointed out by~\cite{Governato2010} and~\cite{Pontzen2012}. 
These authors used zoom-in
simulations of a dwarf galaxy to show that the ``typical'' density at
which the gas becomes eligible to form stars determines the ability of
baryonic processes to transform the central cusp into a core.~\cite{Pontzen2012} showed that if stars
form from relatively low-density gas ($\rho \sim 0.1 \rm \ cm^{-3}$),
the halo remains cuspy because the gas is essentially removed from the
system and its contribution to the gravitational potential never becomes important. If the density threshold for star formation
($\rho_{\rm th}$) is high enough, gas is able to perturb the central
gravitational potential of the halo before turning into stars;
episodic star formation and subsequent supernovae-driven winds then
drive a reduction of the inner mass density of the halo.

\begin{figure}
	\includegraphics[]{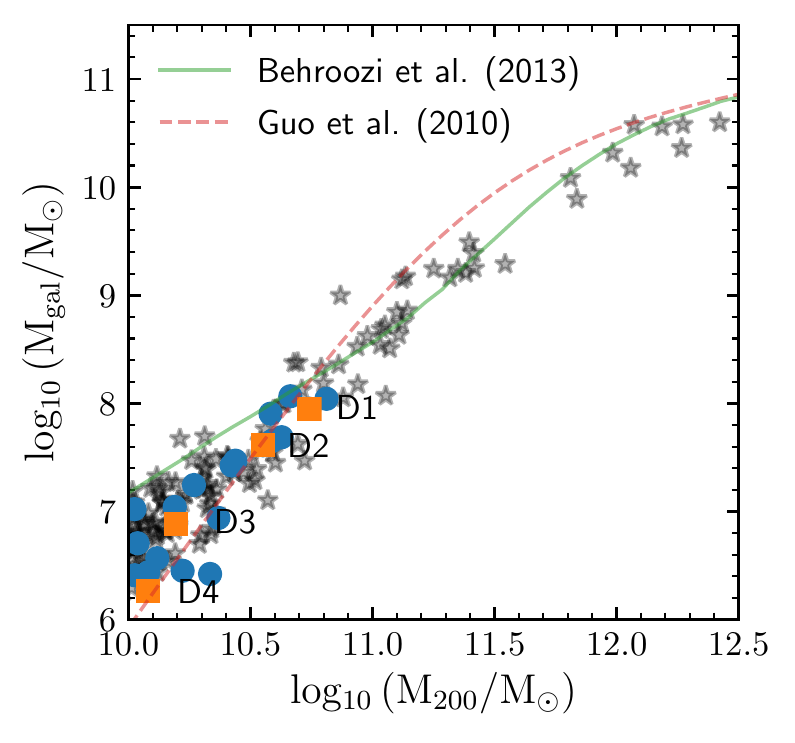}
        \caption{Stellar mass, $M_{\rm gal}$, as a function of halo
          mass, $M_{200}$, for ``central'' galaxies in the parent $12$
          Mpc side-length cosmological hydrodynamical simulation (grey
          stars). Blue solid circles show zoom-in candidate dwarf
          galaxies that fullfil our selection criteria: halo mass in
          the range
          $10^{10} < M_{\rm 200}/M_{\odot} < 10^{11}$ and no
          ``central'' galaxy companion closer than 10 times their virial
          radius. Orange squares show the 4 dwarf galaxies that we
          simulate at higher resolution. Solid and dashed lines show
          abundance matching expectations from
          \protect\cite{Behroozi2013} and \protect\cite{Guo2010}, 
          respectively.}
    \label{Fig:Mgal_M200}
\end{figure}

Although these authors considered only two values of $\rho_{\rm th}$,
their numerical experiments do not only suggest that there must be a
characteristic density threshold for star formation above which the
inner density profile of low-mass haloes is effectively perturbed, but
also that the structural properties of these haloes must inevitably
depend on the value of this parameter. It is then surprising that, to
date, few systematic studies of the impact of this parameter on the
inner density profile of dwarf galaxies have been performed.

In this paper, we explore these ideas further in the context of the
EAGLE model of galaxy formation. We use a set of zoom-in
cosmological simulations of the formation of dwarf galaxies to
demonstrate conclusively that the inner structure of their dark matter
haloes is affected by the choice of the density threshold above which
gas is deemed eligible to form stars. We describe the code and 
simulations in Sec.~\ref{Sec:Simulations}; we show and apply our main
results in Sec.~\ref{Sec:Results} and Sec.~\ref{Sec:EAGLE_cores},
respectively, and end with our conclusions in
Sec.~\ref{Sec:Conclusions}.

\section{The Simulations}
\label{Sec:Simulations}

We first describe the simulation code used in this work and then
provide details of the simulations themselves.

\subsection{The Code} 
\label{subsec:The Code}

\begin{figure*}
	\includegraphics[]{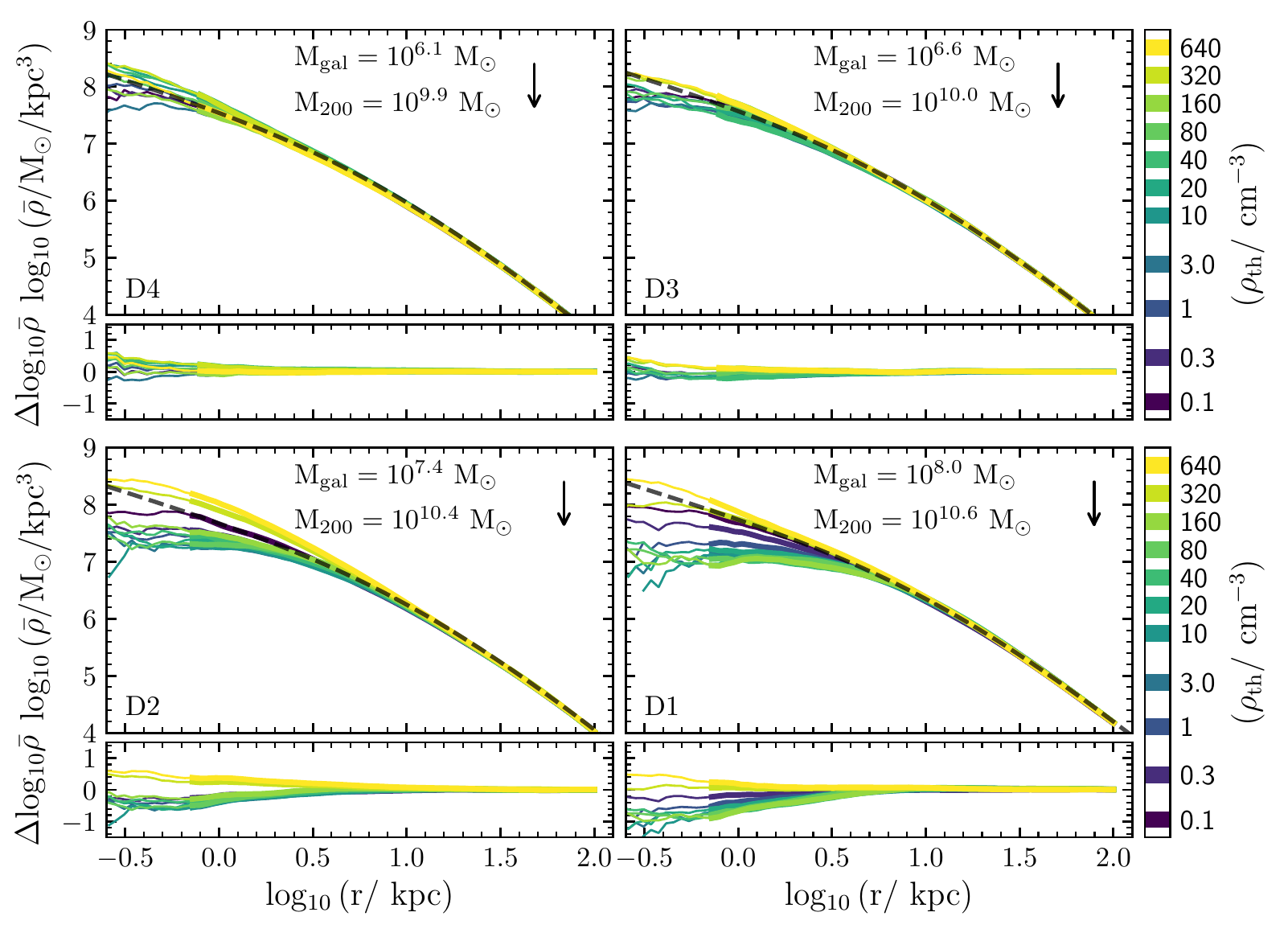}
        \caption{Mean enclosed dark matter density profiles (Eq.~\ref{Eq:density}) at
          redshift $z=0$ of our suite of zoom-in dwarf galaxies,
          simulated at a resolution level L2. The curves show the
          result of varying the assumed gas density threshold for star
          formation, $\rho_{\rm th}$, as indicated in the colour bar
          on the right. The smaller panels show the ratio of the
          individual density profiles to the ``fiducial'' profile of
          the simulation with the lowest value of $\rho_{\rm th}$
          ($0.1 \rm \ cm^{-3}$). The panels are arranged according to
          the galaxy mass, with the least massive example (D4) on
          the top left and the most massive example (D1) on the
          bottom right (see also Fig.~\ref{Fig:Mgal_M200}). Lines
          become thin below 0.5 times the \citealt{Power2003}
          convergence radius, $r_P$. For reference, the dashed
          curves are NFW profiles. For
          $\rho_{\rm th} = 0.1 \rm \ cm^{-3}$, the mean enclosed dark
          matter density profiles are clearly well described by NFW
          profiles for $r> r_{\rm P}/2$. Downward arrows
          indicate the virial radius of the system.}

    \label{Fig:density_profiles}
\end{figure*}

The simulations were performed with the version of the P-Gadget3 code (last described by~\citealt{Springel2005}) used for the 
  EAGLE project~\citep{Schaye2015, Crain2015}. The EAGLE code
includes a set of subgrid prescriptions to model radiative cooling,
photo-heating, star formation and evolution, feedback from star
formation, black hole growth and AGN feedback. The necessary
parameters were originally calibrated by requiring that simulations of
a cosmological volume should approximately match the galaxy stellar
mass function, the distribution of galaxy sizes and the amplitude of
the black hole {\em vs} stellar mass relation.  The original
simulation, as well as its higher resolution counterpart, ``RECAL,''
\citep[see nomenclature in][]{Crain2015} turned out to match not only
these observables, but also other population statistics and scaling
relations as a function of redshift, as well as properties such as
galaxy rotation curves~\citep[e.g.,][]{Schaller2015, Ludlow2017} and the
luminosity function of galactic satellites~\citep[][]{Sawala2016} at
the present day. In this paper we shall refer to the ``RECAL''
calculation as the fiducial EAGLE model.

We summarize here the aspects of the subgrid modelling of star
formation and supernova feedback that are most relevant for this work,
and refer the reader to the original EAGLE papers for a more 
extensive description. 

Star formation proceeds at the Kennicut-Schmidth rate in gas particles
whose density is higher than a threshold, $\rho_{\rm th}$, at
a temperature $T \sim 10^4 \rm \ K$. The EAGLE model assumes a
metallicity-dependent density threshold (as proposed
by~\citealt{Schaye2004}): 
\begin{equation}
\rho_{\rm th} = {\rm min} \left [ \rho_{\rm min} \left (
    \displaystyle\frac{Z}{Z_{0}} \right )^{-\alpha}, \rho_{\rm max}
\right ], 
\end{equation}
where $\rho_{\rm min} = 0.1 \rm \ cm^{-3}$,
$\rho_{\rm max} = 10 \rm \ cm^{-3}$, $Z_0 = 0.002$, and
$\alpha = 0.64$. For simplicity, in this work we depart from this
relation and assume a unique density threshold.  In practice, this is
achieved by setting $\alpha = 0$ and
$\rho_{\rm min} = \rho_{\rm max}$.
The EAGLE code also imposes a temperature floor, $T(\rho) =
T_{0} (\rho / \rho_{\rm th})^{\gamma-1}$, for densities above
$\rho_{\rm th}$, where $\gamma = 4/3$ and $T_{0} = 8000$\,K. We
retain this condition. 

Supernova feedback is implemented following the stochastic thermal
feedback scheme of~\cite{DallaVecchia2012}, namely, newly-formed star
particles inject energy into some fraction of their neighbouring gas
particles. In our simulations - in which the feedback implementation
is identical to that of \cite{Schaye2015} and \cite{Crain2015} - the
energy of each supernova explosion is deposited, on average, in one
gas particle near each evolved stellar particle. The energy
injection is implemented by heating gas particles to a temperature of
$10^{7.5} K$, without turning radiative cooling off. Feedback
particles thus receive an amount of thermal energy equal to:
\begin{equation}
\Delta E = \displaystyle\frac{\left (\gamma-1 \right ) }{\mu}  k_{B}
\Delta T \left ( \displaystyle\frac{m_{g}}{m_{p}} \right ), 
\end{equation} 
\noindent where $\Delta T = 10^{7.5}$\,K. This ensures that, by construction, the gas cooling time of feedback particles is sufficiently long that supernovae energy can couple efficiently to the interstellar medium.

\subsection{Zoom-in dwarfs}
\label{Sec:Zoom-in}

Our systematic study is based on high-resolution zoom-in cosmological
hydrodynamical simulations of the formation of isolated dwarf galaxies
with a relatively wide range of halo virial\footnote{Virial quantities
  correspond to those of the sphere within which the enclosed mean
  mass density is 200 times the critical density of the Universe,
  $\rho_{\rm crit} = 3 H_{0}^2/8\pi G = 2.775 \times 10^{11} h^2
  M_{\odot} / \rm \ Mpc^3$, and are identified with the subscript,
  $200$. Throughout this paper, we assume $h = 0.704$.} mass
($10^{10} < M_{\rm 200}/M_{\odot} < 10^{11}$). Initial conditions are
obtained from the publicly available code,
  MUSIC~\citep{Hahn2011}, which generates zoom-in initial conditions
by refining a region centred around an object of interest selected
from a parent cosmological volume, while degrading the resolution
farther out. Our assumed cosmological parameters are consistent with
the WMAP-7 values~\citep{Komatsu2011}: $\Omega_{m} = 0.272$,
$\Omega_b = 0.0455$ and $\Omega_{\Lambda} = 0.728$, $h = 0.704$,
$\sigma_8 = 0.81$ and $n_{s} = 0.967$\footnote{We adopt the WMAP-7
  cosmological parameters rather than the Plank~\citep{Planck2016}
  values in order to be consistent with the parameters adopted in the
  APOSTLE project. The choice of cosmological parameters has a
  negligible effect on the small scales studied in this work.}.

Dwarf galaxies were identified in a parent cosmological periodic volume of side 12~Mpc simulated at a resolution level L1, using the EAGLE fiducial model. The volume is filled with $256^3$ dark
matter particles and the same number of gas particles, so that the
dark matter particle mass is
$m_{\rm drk} = 3.2 \times 10^6 \ M_{\odot}$, and the gas particle mass
is $m_{\rm gas} = f_{b} \ m_{\rm drk} = 5.3 \times 10^5 \ M_{\odot}$,
where $f_{b} = \Omega_{b}/\Omega_{m}$ is the universal baryon
fraction. The identification of dwarf candidates was performed at
redshift $z=0$ using the group finder SUBFIND~\citep[][]{Springel2001, Dolag2009}, which identifies
self-bound substructures within a catalogue of friend-of-friends (FoF)
haloes built using a percolation length of 0.2 times the mean
interparticle separation~\citep{Davis1985}. SUBFIND produces a
catalogue of ``central'' and ``satellite'' subhaloes within each FoF
halo. Our analysis is based on ``central'' galaxies only for which, in
addition, we apply an ``isolation'' criterion by requiring that they
should have no ``central'' companion closer than 10 times their virial radius.

Our mass and isolation criteria yield a sample of 21 dwarf candidates,
from which we select 4 that sample our chosen ranges of halo
($10^{10} < M_{\rm 200}/M_{\odot} < 10^{11}$) and stellar
($10^{6} < M_{\rm gal}/M_{\odot} < 10^{8}$) masses. Our simulation
suite thus consists of 4 zoom-in cosmological simulations of the
formation of isolated dwarf galaxies. These were all
simulated at resolution level, L2, in which the gas particle mass is
$m_{\rm gas} = 6.6 \times 10^4 \ M_{\odot}$ and the dark matter
particle mass is $m_{\rm drk} = 3.9 \times 10^5 \ M_{\odot}$. 

The zoom-in simulations were performed keeping all the parameters of
the EAGLE fiducial model fixed, as explained in Sec.~\ref{subsec:The
  Code}, but systematically varying the density threshold for star
formation, $\rho_{\rm th}$, from the lowest value considered in the
EAGLE fiducial model, $\rho_{\rm th} = 0.1 \rm \ cm^{-3}$, to a
largest value, $\rho_{\rm th} = 640 \rm \ cm^{-3}$. The latter is
slightly lower than the values adopted in simulations that produce
cores in dwarf galaxies, such as FIRE-2~\citep[e.g.,][]{Fitts2017}. Finally, the gravitational
softening of the dark matter, gas and stellar particles is chosen so
that its value never exceeds $1\%$ of the mean interparticle
separation. This yields $\epsilon \sim 500 \rm \ pc $ for the parent
volume (resolution level~L1), and $\epsilon \sim 234 \rm \ pc$ for the
zoom-in dwarfs (resolution level~L2).

Fig.~\ref{Fig:Mgal_M200} shows the stellar mass, $M_{\rm gal}$, as a
function of the virial mass, $M_{200}$, for ``central'' galaxies
in the parent volume. Grey stars show all luminous
galaxies\footnote{more than 1 stellar particle within a sphere of
  radius, $R_{200}$.} in the volume down to a virial mass,
$M_{200} \sim 10^{10} \ M_{\odot}$. The blue solid circles mark the 21
galaxies that fulfil our selection criteria of being isolated dwarfs
in haloes of virial mass $10^{10} < M_{\rm 200}/M_{\odot} <
10^{11}$.
The orange squares are the 4 zoom-in dwarf galaxies that we shall use
for further analysis, and which span the entire range of halo and
stellar mass of interest. The high-mass galaxies are consistent with
abundance matching expectations from~\cite{Behroozi2013} (solid line)
whereas, as shown by \cite{Sawala2015}, the low-mass galaxies
already begin to deviate from these expectations and, for the mass
range plotted, lie between the extrapolated abundance matching
relations of \cite{Behroozi2013} and~\cite{Guo2010} (dashed line).
 
\section{Results}
\label{Sec:Results}

\subsection{Role of the star formation density threshold}

Our main result is shown in Fig.~\ref{Fig:density_profiles}, where we
plot the radial dependence of the mean enclosed dark matter density
profile,
\begin{equation}
\label{Eq:density}
\bar \rho(r) = \displaystyle\frac{M(r)}{4/3 \pi r^3}
  = \displaystyle\frac{3}{r^3} \displaystyle\int_{0}^{r} \rho(r') r'^2
  \ dr',
\end{equation}
for our four (L2) zoom-in dwarf galaxies. Curves are coloured according to the value of the gas density threshold for star formation, as indicated by the colour bar on the right. The value of
$\rho_{\rm th}$ varies from $0.1 \rm \ cm^{-3}$ (hereafter LT) to an arbitrary high value, $640 \rm \ cm^{-3}$ (HT). The LT simulations are {\it all} consistent with NFW profiles (dashed lines) above $\sim 0.5$ times the convergence radius, $r_P$, defined by~\citet[][]{Power2003}. (Below $r_P$, the density profiles of the LT runs are slightly shallower than the
reference NFW profile as a result of numerical relaxation.) For higher
values of $\rho_{\rm th}$, the dark matter density profiles, in some cases, depart systematically from NFW. The subpanels show
the ratio of the individual profiles to the profile of the corresponding LT run. Clearly, the assumed value of $\rho_{\rm th}$ has little impact
on the inner dark matter density profiles of dwarfs D4 and D3, but
dramatically alters those of dwarfs D2 and D1.

\begin{figure}
	\includegraphics[]{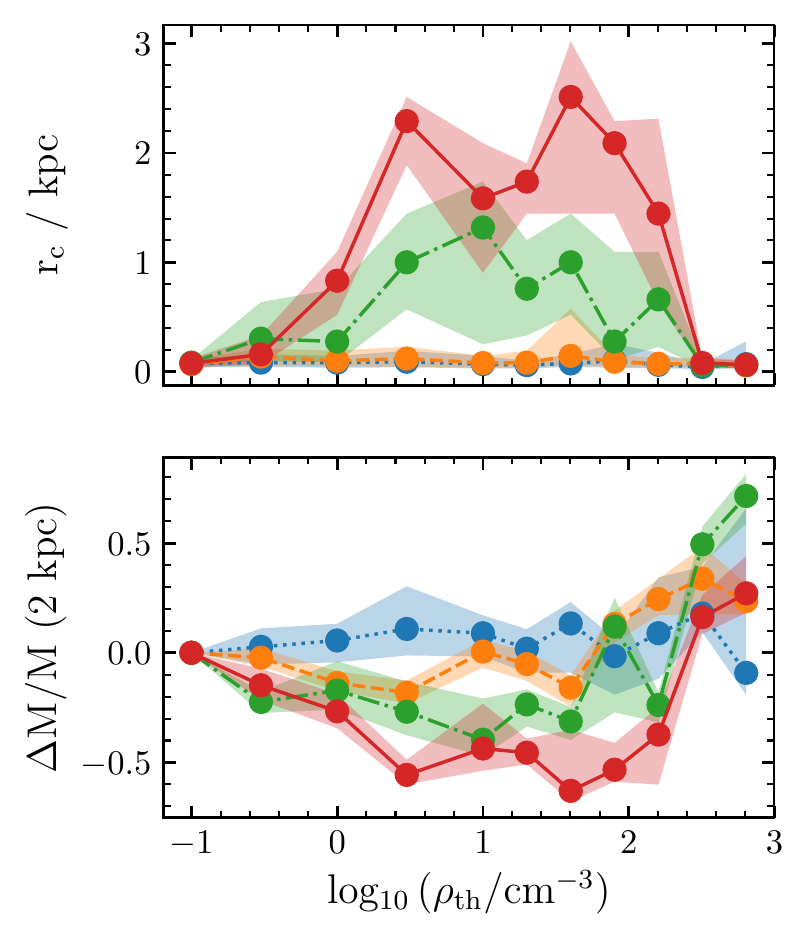}
        \caption{The top panel shows the median core radius, $r_c$, 
          as a function of the
          assumed gas density threshold for star formation,
          $\rho_{\rm th}$. The different simulated dwarfs are
          identified by different line types and colours, as labelled.           
          $r_c$ is
          defined as the radius at which the mean enclosed dark matter
          density of the halo drops by a factor of 2 relative to the
          LT ($\rho_{\rm th} = 0.1 \rm \ cm^{-3}$) simulation. If
          $r_c$ does not exist because the halo density profile is
          either similar or steeper than that of the fiducial
          simulation, we set $r_{c} = 0$. The bottom panel shows the 
          median dark
          matter mass deficit within the inner $2 \rm \ kpc$ relative
          to the fiducial LT simulation. Medians are taken over time, using 100 snapshots equally spaced in time, after redshift $z=1$. Shaded regions show the 10th-90th percentiles of the distributions.}
    \label{Fig:core_radius} 
\end{figure}

Consider the mean enclosed density profile of D1 (bottom-right
panel) at $1 \rm \ kpc$: it decreases by a factor of $\sim 3$ as
$\rho_{\rm th}$ increases from $0.1 \rm \ cm^{-3}$ to
$1 \rm \ cm^{-3}$, and by a factor of $\sim 5$ when
$\rho_{\rm th} \sim 100 \rm \ cm^{-3}$.  For
$\rho_{\rm th} = 320 \rm \ cm^{-3}$ the profile is now very similar
to that of the LT case and, for higher values of $\rho_{\rm th}$, it
becomes slightly denser. Thus, for the wide range of values of
$\rho_{\rm th}$ considered here
($0.1<\rho_{\rm th} / \rm cm^{-3} <640$), the inner dark matter
density of D1 at first decreases roughly monotonically with
increasing $\rho_{\rm th}$, but then increases for higher values of
the threshold. D2 exhibits a similar behaviour.

We show this quantitatively in Fig.~\ref{Fig:core_radius}. The
top panel shows the median ``core'' radius, $r_c$, as a function of
$\rho_{\rm th}$ for the four simulated dwarf galaxies. We define the
``core'' radius as the radius within which the mean enclosed dark
matter density of the halo drops by a factor of 2 relative to the LT
case \footnote{Note that, by construction, this is also the radius at
  which the enclosed dark matter mass is lower by a factor of 2 than
  in the fiducial LT case. We assume $r_{c} = 0$ if the core radius
  does not exist for that system.}. Medians are taken over time using 100 snapshots equally spaced in time after redshift $z=1$. This demonstrates that, as anticipated in Fig.~\ref{Fig:density_profiles}, $r_{c}$ is effectively a function of $\rho_{\rm th}$ for the more massive dwarfs D1 and D2, but the value of $\rho_{\rm th}$ has little impact on the inner structure of dark matter haloes of the less massive dwarfs D3 and D4. The dependence of $r_c$ on halo mass, which holds over a wide range of $\rho_{\rm th}$ ($3 \le \rho_{\rm th}/ {\rm cm^{-3}} \le 160$), is broadly consistent  with earlier work that showed that very low-mass dwarfs do not exhibit large dark matter cores~\citep[][and references therein]{DiCintio2014b, Tollet2016, Bullock2017}. However, the scatter in $r_c$ over time is large (the shaded regions show the 10th-90th percentiles of the temporal distribution of $r_c$, after $z=1$). We will show in Sec. 3.3 that these dwarfs can experience a number of cusp/core transformations over their lifetime.

Consider again galaxy D1, which has a cuspy halo in the EAGLE
fiducial model (LT run). This dwarf develops a core of
$r_{c} \sim 1 \rm \ kpc$ when simulated assuming
$\rho_{\rm th} \sim 1 \rm \ cm^{-3}$, and a core of
$r_{c} \sim 2 \rm \ kpc$ when simulated assuming
$\rho_{\rm th} \sim 10 \rm \ cm^{-3}$. This demonstrates that the
particular choice of the gas density threshold at which stars are
allowed to form in a simulation determines not only the creation (or
not) of a core, but also that the size of the core itself (or the
amount of mass displaced from the central regions) depends on the
exact choice of threshold. This is consistent with the results of~\cite{Governato2010} and~\cite{Pontzen2012, Pontzen2014}, who probed only two values of $\rho_{\rm th}$ ($0.1-100 \rm \ cm^{-3}$).

The bottom panel of Fig.~\ref{Fig:core_radius} shows
$\Delta M/M(2 \rm kpc)$, the dark matter mass deficit (or excess if
$\Delta M >0$) in the inner $2 \rm \ kpc$ relative to the LT
simulation, as a function of $\rho_{\rm th}$. We can see that the dark
matter mass enclosed within this (arbitrary) radius is, as $r_c$ is
too, very sensitive to the choice of $\rho_{\rm th}$ for most dwarfs,
and can be significantly reduced for intermediate values of
$\rho_{\rm th}$. For very high values of the threshold
($\rho_{\rm th} > 160 \rm \ cm^{-3}$), some dwarfs develop an excess of
dark matter mass within $2 \rm \ kpc$, not a deficit, relative
to their LT counterpart, as anticipated from
Fig.~\ref{Fig:density_profiles}. We now explore the reasons why the
core radius -- or the mean dark matter mass enclosed by an inner
radius -- is so sensitive to the assumed density threshold for star
formation.

\begin{figure*}
	\includegraphics[]{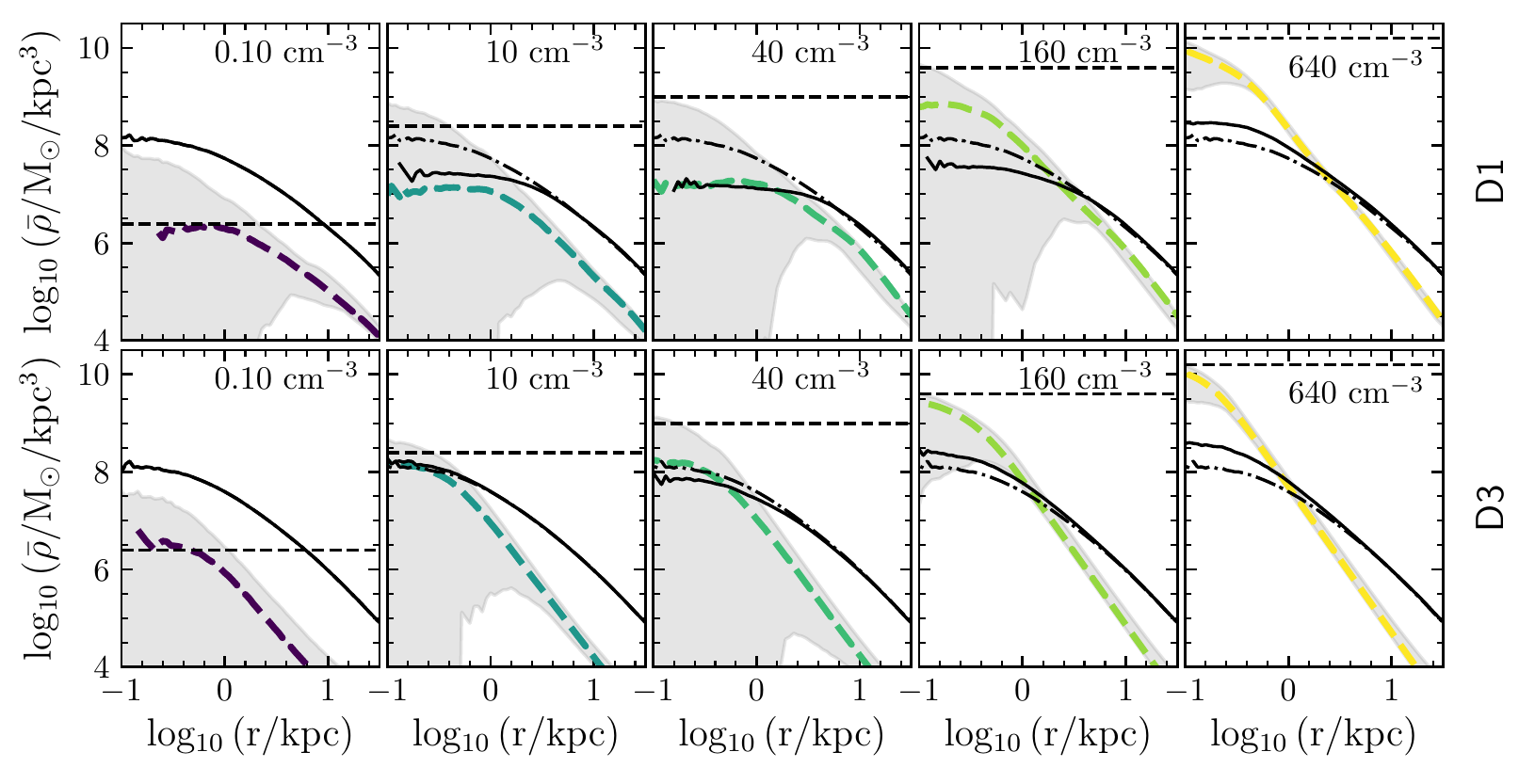}
        \caption{The mean enclosed dark matter density profile (solid
          black lines) of dwarfs D1 (top panels) and D3 (bottom
          panels) simulated assuming different density thresholds for
          star formation (shown by the horizontal dashed line and 
          listed in the legend of each panel).  The
          dot-dashed curves show the mean enclosed dark matter density
          profile measured in the EAGLE fiducial run. The colour
          dashed curves show the median of the enclosed gas
          density profile. Medians are taken over time after the halo has
          collapsed and the overall dark matter density profile within
          $r \sim 3 \rm \ kpc$ has stabilized (typically for time
          $t>3.5 \, \rm Gyr$, Fig.~\ref{Fig:origin_cores_all}).  The
          shaded regions show, at a given radius, the minimum and
          maximum values that the gas profile ever reached during this period.}
    \label{Fig:density_profiles_D1}
\end{figure*}

\subsection{The impact of the density threshold for star formation on the gas density profile of low-mass galaxies}

The top panels of Fig.~\ref{Fig:density_profiles_D1} show the median
of the enclosed dark matter density profile of simulated dwarf
D1 (black solid lines), for different values of the gas density
threshold for star formation. The median is taken over a time
interval starting well after the halo has collapsed and its overall
dark matter density profile has stopped evolving ($t \gtrsim 3.5 \rm \ Gyr$)
and ending at the present day. We also show the median of the enclosed gas
density profiles (coloured dashed lines), and the minimum and maximum
mean enclosed gas density at each radius (shaded regions), measured
also for $t \gtrsim 3.5 \rm \ Gyr$)\footnote{We performed the
  measurements over 150 snapshots equally spaced in time over the range
  $3.5 \rm \ Gyr \lesssim t \lesssim 13.7 \rm \ Gyr$.}. This figure
provides us with some immediate insight into why $r_c$
depends on $\rho_{\rm th}$.

Consider, for example, the fiducial LT simulation
($\rho_{\rm th} = 0.1 \rm \ cm^{-3}$). In this case, the dark matter
halo remains cuspy during its entire lifetime, including at $z=0$.
The reason for this is straightforward: although supernova feedback is
effective at removing baryons from the central region of the halo (as
discussed in Sec.~\ref{Sec:Introduction}, this is a necessary
condition for reducing the inner dark matter density), gas never
contributes significantly to the gravitational potential of the
system, which is also a necessary condition for baryons to perturb the mass
profile of the halo. Indeed, gas is either turned into stars or
expelled from the inner regions before reaching sufficiently high
densities. The low gas density threshold for star formation,
$\rho_{\rm th} = 0.1 \rm \ cm^{-3} \sim 1.8 \times 10^{6} \rm \
M_{\odot} / \ kpc^{-3}$, ensures that little gas can become denser
than this value; the median gas density profile, shown by the colour
dashed line remains under this threshold (indicated by the black
horizontal dashed line) at all radii.

Consider now a simulation of D1 with a higher value of the gas density
threshold for star formation, say $\rho_{\rm th} = 10 \rm \
cm^{-3}$. The qualitative evolution of the galaxy is similar to that
of its LT counterpart, but there is a significant quantitative
difference: more gas collects in the inner region of the halo before
turning into stars. The gas density threshold for star formation, now
$100$ times larger than in the LT simulation, is much higher than any
resolved dark matter density. The consequence is clear: baryons sometimes
dominate the gravitational potential of the system in the inner
$\sim 2 \rm \ kpc$ before turning into stars or being expelled by
feedback processes. Thus, in contrast to the LT example, baryons can
now readily escape from the central regions of the system, but only
after their gravitational pull has affected the dark matter
halo. These are indeed the two ingredients commonly invoked in the
literature that can lower the inner density of dark matter
haloes~\cite[][]{Navarro1996b,Governato2010,Pontzen2012}. The resulting $z=0$
dark matter density profile (solid line) is, in fact, significantly
shallower than its LT counterpart in the inner $\sim 2 \rm \ kpc$. A
similar behaviour is observed for higher density thresholds (e.g.,
$\rho_{\rm th} \sim 40 \rm \ cm^{-3}$), although in this case, the
extent of the region in which the dark matter density profile is
affected is larger, as anticipated from Fig.~\ref{Fig:core_radius}.

\begin{figure}
	\includegraphics[]{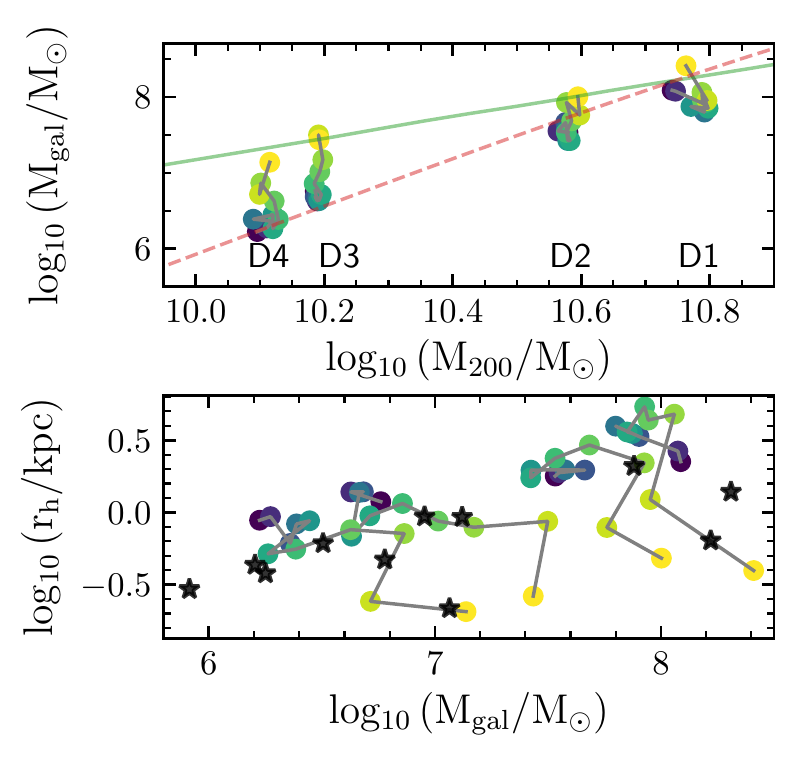}
        \caption{Top panel: the stellar mass of our suite of simulated
          dwarfs as a function of halo mass, for different values of
          the gas density threshold for star formation,
          $\rho_{\rm th}$ (colour coding is as in
          Fig.~\ref{Fig:density_profiles}). Points that
          correspond to the same galaxy are joined by lines.  
          The red dashed and green lines show extrapolations of the \protect\cite{Guo2010} and \protect\cite{Behroozi2013} abundance matching relations, respectively. Bottom
          panel: the half-mass radius of the stellar component as a
          function of galaxy mass. Starred symbols show observations
          of field dwarf galaxies in the LG, compiled by
          \protect\cite{Fattahi2018}.
          The stellar mass and half-mass radius barely change for
          low/moderate values of $\rho_{\rm th}$; however, when increasing
          this threshold beyond
          $\rho_{\rm th} \gtrsim 80 - 160 \rm \ cm^{-3}$, the stellar
          masses typically increase by more than a factor of 5, with
          the result that the galaxies become more centrally
          concentrated.}
    \label{Fig:Mgal_radius_rho}
\end{figure}

What determines the size of the core radius? Using our previous
insight, the core radius must be
related to the extent of the region where the gravitational force of
the gas is non-negligible compared to that of the dark matter
halo. For a spherically-symmetric system, this would occur at the
radius where the mean enclosed gas density profile becomes comparable
to that of the dark matter\footnote{This is the case because, at a
  given radius,
  $\bar \rho_{\rm dm}/\bar \rho_{\rm gas} = \rm M_{\rm dm}(<r) /
  M_{\rm gas}(<r)$.}. The top panel of
Fig.~\ref{Fig:density_profiles_D1} indicates that this is the case
for most values of $\rho_{\rm th}$, but not for very
high gas density thresholds for star formation
($\rho_{\rm th} = 320 \rm \ cm^{-3}$ and
$\rho_{\rm th} = 640 \rm \ cm^{-3}$), for which $r_{c} = 0$. For these
values of $\rho_{\rm th}$, the gas accumulates in the centre and
becomes gravitationally dominant before turning into stars but is
unable to evacuate the central regions efficiently because of the
enhanced potential well depth caused by the accumulation of baryons,
and because the EAGLE feedback scheme becomes inefficient at such high densities.

Consider, for example, the simulation with the highest gas density
threshold for star formation, $\rho_{\rm th} = 640 \rm \ cm^{-3}$
(rightmost top panel of
Fig.~\ref{Fig:density_profiles_D1}). Although the gas density
profile fluctuates in the innermost regions of the halo (see shaded
region), the fluctuations are so small that the system remains
baryon-dominated at all times and the dark matter halo contracts
significantly in response. In principle, one might suspect
that this could be due to lack of resolution to follow the collapse of
high-density gas clouds and therefore, star formation events. However,
we shall demonstrate below that this is not the case.

Fig.~\ref{Fig:Mgal_radius_rho} summarises the changes in the main
structural properties of our four simulated dwarf galaxies as the
star-formation threshold, $\rho_{\rm th}$, varies. (The colour coding
is as in Fig.~\ref{Fig:density_profiles}). The top panel is analogous
to Fig.~\ref{Fig:Mgal_M200} and shows the stellar mass - halo mass
relation of the simulated dwarfs for different values of
$\rho_{\rm th}$. Not surprisingly, for low values of $\rho_{\rm th}$
(black dots), the stellar mass of the dwarfs is essentially consistent
with the stellar mass they have in the parent volume from which they
were selected (orange squares in Fig.~\ref{Fig:Mgal_M200}). For a wide
range of values of $\rho_{\rm th}$, the stellar mass changes
little. However, it increases significantly, by more than a factor of
$3$, for very high values of $\rho_{\rm th} > 160 \rm \
cm^{-3}$. This plot clearly demonstrates that, for the highest values 
of $\rho_{\rm th}$, our simulated galaxies form even more stars than
their LT counterparts.

Note, however, that the stellar content of the galaxies, regardless of
$\rho_{\rm th}$, never accounts for more than $0.2 \%$ of the total
mass of the system, or for more than $5 \%$ of the mass within
$2 \rm \ kpc$. Thus, for large values of $\rho_{\rm th}$, most of the
baryonic component of the galaxy is kept at the very centre in the
form of gas, reflecting a significant reduction in the
efficiency of feedback to drive gas out from high-density
regions, which deepens the central gravitational potential. Indeed, the minimum requirement for supernovae explosions to
push gas out to large radii is that the sound-crossing time,
$t_{sc}$, of the expanding gas shell be much smaller than the cooling
time, $t_{c}$. For the EAGLE feedback implementation $t_{sc}$
becomes larger than $t_{c}$ for densities higher than
$\rho_{\rm th} \sim 30 \rm \ cm^{-3}$~\citep[see Eq. 18
of][]{DallaVecchia2012}, which is close to the density threshold for
star formation that produces the largest cores in our simulations (see
e.g., Fig.~\ref{Fig:core_radius}). It is then not unexpected that a
significant contraction of the dark matter halo is observed for high
values of $\rho_{\rm th}$.

The bottom panel of Fig.~\ref{Fig:Mgal_radius_rho} shows the
half-stellar-mass radii of the simulated dwarfs as a function of their
stellar mass, for different values of $\rho_{\rm th}$. The black stars
show data for ``field'' dwarfs in the LG compiled
by~\cite{Fattahi2018}. The value of $\rho_{\rm th}$ has a significant
impact on the size of the most massive dwarfs. The reduction of the inner
dark matter density produced by gas blowouts also causes the galaxies
themselves to expand~\citep[see e.g.,][for a similar result]{Dutton2016, ElBadry2016}. Note, however, that the effect is mild, and non-monotonic. Indeed, increasing the density threshold to very high values makes the galaxies not only more massive but also more compact.

\begin{figure*}
	\includegraphics[]{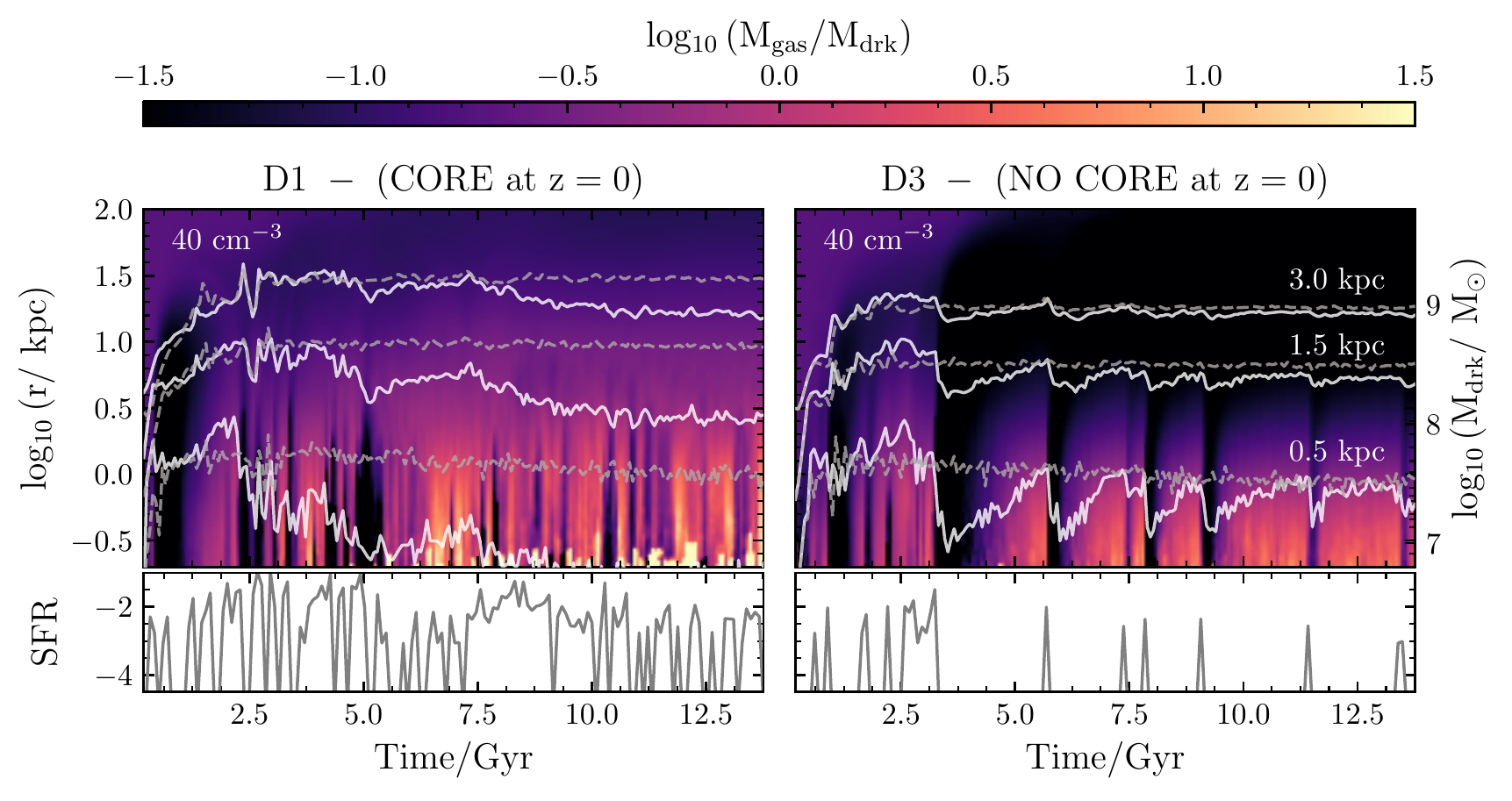}
        \caption{Enclosed gas-to-dark matter mass ratio (colour scale
          at the top), as a function of distance to the main
          progenitor (y-axis) and time (x-axis). The left and right columns
          correspond to the simulated dwarfs D1 and D3, respectively. In both cases, $\rho_{\rm th} =40 \rm \ cm^{-3}$. The
          solid lines show the enclosed dark matter mass (scale on the
          right) at three different radii: $0.5 \rm \ kpc$ (bottom
          line), $1.5 \rm \ kpc$ (middle line) and $r=3 \rm \ kpc$
          (upper line). The dashed lines are these same quantities,
          but for the $\rho_{\rm th} = 0.1 \rm \ cm^{-3}$ (LT)
          simulation. The lower panels show the decimal logarithm of the star formation rate of the galaxies, in units of $M_{\odot} / \rm \ yr $, measured in bins equally spaced by $\sim 100 \rm \ Myr$. }
    \label{Fig:origin_cores_all} 
\end{figure*}

A robust prediction of the value of $r_c$ for a given value of
$\rho_{\rm th}$ is difficult: the core radius depends not only on the
radius within which the gas becomes important at sourcing the
gravitational potential but also on the radius above which feedback
processes, such as supernovae explosions, are effective at pushing gas
out. This will ultimately depend on the feedback implementation of
each particular model. This also implies that idealised models used to calculate core radii~\cite[e.g.][]{Penarubia2012,Maxwell2015}, are expected to have limited applicability~\cite[see also][]{Pontzen2015}. Indeed, dwarfs D1 and D2 exhibit very different dark matter profiles when simulated with, e.g., $\rho_{\rm th} = 0.1 \rm \ cm^{-3}$ and $\rho_{\rm th} = 10 \rm \ cm^{-3}$, yet they form virtually the same amount of stars, implying that they have been subject to similar feedback processes over their lifetime. The difference in their dark matter profiles is mostly driven by the gravitational coupling between the gas and the dark matter halo, a crucial ingredient usually not taken into account in analytic models. 

Finally, taken at face value, the LG data in
Fig.~\ref{Fig:Mgal_radius_rho} are broadly consistent with
intermediate values of $\rho_{\rm th} \sim (10-100) \rm \ cm^{-3}$, but given the uncertainties just mentioned, we cannot confidently rule out any of the values of $\rho_{\rm th}$ considered in this work.

We now focus on D3, whose dark matter density profile is largely
insensitive to the choice of $\rho_{\rm th}$ (see top right panel of
Fig.~\ref{Fig:density_profiles} or dashed line in
Fig.~\ref{Fig:core_radius}). The bottom panels of
Fig.~\ref{Fig:density_profiles_D1} show that for low values of
$\rho_{\rm th}$, the baryons are unable to perturb the inner dark
matter density profile, just as in D1 illustrated in the top left panels
of the figure. For the highest value of $\rho_{\rm th}$, the dark
matter halo becomes more centrally concentrated than in the LT
simulation due to a significant increase in the gas-dominated
gravitational potential. However, for intermediate values of $\rho_{\rm th}$, no significant reduction in the inner dark matter density profile is observed at redshift $z=0$. We now explore the reasons why simulated dwarf D3, which appears to satisfy
the conditions required for baryonic processes to reduce its central
dark matter density profile for a wide range of $\rho_{\rm th}$,
remains largely undisturbed.

\subsection{Star formation history of dwarfs and its impact on their inner dark matter content}

For a large range of values of $\rho_{\rm th}$ neither D3 nor D4
experience a significant reduction in their inner dark matter
density, despite: (i) having been baryon dominated in the past and
(ii) having undergone significant baryonic blowouts.  In order to
understand this behaviour we explore the evolution and assembly
history of D3 and contrast it with that of D1 which does end up with a
reduced inner dark matter density.

Fig.~\ref{Fig:origin_cores_all} shows the enclosed gas-to-dark matter
mass ratio (colour scale on the top) as a function of time (x-axis)
and distance to the centre of the main progenitor (y-axis) for D1 and
D3 and $\rho_{\rm th}=40 \rm \ cm^{-3}$. From top to bottom in each
panel, the solid lines show the enclosed dark matter mass within
$3.0 \rm \ kpc$, $1.5\rm \ kpc$, and $0.5 \rm \ kpc$,
respectively. The dashed lines show the same quantities in the LT run
(i.e., for $\rho_{\rm th} = 0.1 \rm \ cm^{-3}$). Individual star
formation histories are shown in the lower panels. Bursts of star
formation precede gas blowouts which are manifest as sudden reductions
of the $M_{\rm gas}/M_{\rm drk}$ ratio.

The collapse of the dark matter haloes of both dwarfs is complete
after a couple of Gyrs, after which the enclosed dark matter mass
within $3 \rm \ kpc$ (and therefore the density within that radius)
remains roughly constant and is almost indistinguishable from that of
the LT counterpart.

Soon after collapsing, dwarf D1 becomes gas dominated
($M_{\rm gas}/M_{\rm drk} > 1$) in the inner $2 \rm \ kpc$ and starts
forming stars in earnest. The first major episode of star formation
occurs at $t \sim 2.5 \rm \ Gyr$ and is accompanied by a significant
reduction of the inner gas mass (see darker regions in the left panel
of Fig.~\ref{Fig:origin_cores_all}). The central gravitational
potential of D1 is thus suddenly reduced, allowing dark matter
particles to migrate to larger orbits, and this causes a reduction of
more than a factor of $3$ in the enclosed mass at $r=0.5 \rm \
kpc$. At $t\sim 5 \rm \ Gyr$, another intense episode of star
formation empties the gas from the inner regions of the halo and also
causes a major loss of dark matter mass. Some gas returns soon after,
but the halo barely changes.

Most of the reduction of the inner dark matter mass takes place in one
or two distinctive blowouts, which empty the central regions during a
few million years. Following these events, the baryon mass at the
centre is replenished but it is removed almost immediately by short
bursts of star formation. Thus, the rapid assembly and removal of
baryons from the central region is less effective at evacuating the 
inner dark matter mass, and therefore at forming cores,
than a single massive blowout.  Consider, for example, the enclosed
dark matter mass at $r=1.5 \rm \ kpc$ (middle line). This drops by
roughly a factor of $2$ after the sudden removal of gas at
$t \sim 5 \rm \ Gyr$, and by a similar amount from $t=7.5 \rm \ Gyr$
to $t=12.5$ as a result of the ``burstiness'' of the system.  We
demonstrate in Appendix~\ref{Sec:App} that the rapid assembly and
removal of gas, although inefficient at forming dark matter cores, can
allow star formation to continue while preventing the halo from contracting.

In the case of D3 (right panel of Fig.~\ref{Fig:origin_cores_all}),
the first important event of star formation takes place at
$t \sim 2.5 \rm \ Gyr$. The associated supernovae explosions violently
expel the central gas from the system and this causes a dramatic
reduction (by a factor of 10) of the dark matter mass within
$r<0.5 \rm \ kpc$, not dissimilar to the case of D1.  After a
protracted period of time ($\Delta t \sim 2 \rm \ Gyr$), more baryons have
cooled and sunk to the centre of the halo, dragging dark matter from
larger radii and allowing the halo to recover most of the inner dark
matter mass that it had had initially. The dark
matter halo, however, does not return to its initial
(contracted) configuration, but to a configuration with a slightly
shallower slope.  At $t \sim 5.5 \rm \ Gyr$ the central gas density in
D3 has become comparable to $\rho_{\rm th}$ and the galaxy undergoes
another episode of star formation, which blows out the gas
again and reduces once more the inner dark matter density. This
process of star formation, gas blowout and recovery of the inner dark
matter density is repeated at least 4 more times before redshift $z=0$.
The absence of a major event of star formation at late times leaves
the dark matter halo of D3 in a configuration that is essentially
indistinguishable from that of the fiducial LT counterpart.

As shown in Appendix~\ref{Sec:AppB}, dwarf D4 exhibits a similar evolution: its dark matter
halo is perturbed by sparse episodes of star formation but recovers, and becomes even more compact than its LT counterpart, as
more baryons condense at the centre. The evolution of D3 demonstrates
that the sudden removal of gas from the centre is very efficient at
reducing the enclosed dark matter mass; however, the lack of
subsequent star formation activity allows gas to return and gradually
accumulate at the centre causing the halo to contract.

The main difference between the evolutionary paths of D1 and D3
following the first episode of core formation stems from the different
timescales on which the collapse of the gas that fuels star formation
occurs in each case. If gas can condense and form stars but is
expelled on a short timescale - as in D1 - the dark matter halo has no
time to respond dynamically and contract and so it retains a shallow
density profile. But if the accretion of gas occurs over a long
timescale - as in D3 - the dark matter halo has time to contract in
response to the accumulated gas at its
centre. Fig.~\ref{Fig:origin_cores_all} indicates that D1 undergoes
continuous perturbations whilst D3 does not. This suggests that the
occurrence of repeated perturbations, although individually
inefficient at perturbing the dark matter halo, serves to prevent the
recovery of the cusp in the halo of D1.

\begin{figure}
	\includegraphics[]{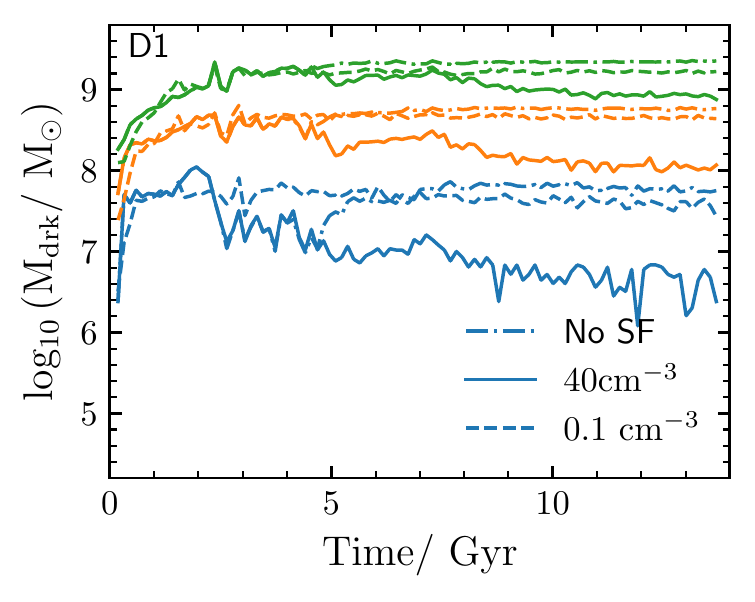}
        \caption{Enclosed dark matter mass within $0.5 \rm \ kpc$,
          $1.5 \rm \ kpc$ and $3 \rm \ kpc$ for simulated dwarf
          D1. From bottom to top different colour lines indicate the
          evolution of the mass within these radii. The dashed lines
          show the result of running the simulation with a density
          threshold of $0.1 \rm \ cm^{-3}$, and the solid lines with a
          threshold of $ 40 \rm \ cm^{-3}$. The dot-dashed lines
          correspond to a simulation identical to that shown by the
          solid lines but in which gas is prevented from turning into
          stars after $t \sim 5 \rm \ Gyr$. This plot demonstrates
          that, for dwarf D1, the central mass is prevented from
          reassembling at the centre by the intense star formation
          activity and accompanying gas blowout.}
    \label{Fig:contration_of_halo1}
\end{figure}

We test this interpretation with a simple experiment. We re-run the
dwarf D1 simulation (assuming a value of
$\rho_{\rm th} = 40 \rm \ cm^{-3}$) until $t \sim 5 \rm \ Gyr$, after
which we artificially halt star formation. Gas is then free to cool
and sink to the centre of the halo. If it is repeated gaseous blowouts
that maintain a shallow central density profile in the dark matter,
then the absence of star formation should allow the halo to contract
rapidly as gas accumulates at its centre. As shown in
Fig.~\ref{Fig:contration_of_halo1} this is indeed what happens. The
solid and dashed lines show the enclosed dark matter mass within 3
different radii, for the fiducial model
($\rho_{\rm th} = 0.1 \rm \ cm^{-3}$) and for the model with
$\rho_{\rm th} = 40 \rm \ cm^{-3}$, respectively; the dot-dashed line
shows the enclosed dark matter mass of the simulation in which star
formation is turned off after $t \sim 5$. Clearly, in the absence of
star formation, the dark matter halo reacts to the collapse of gas by
contracting, with the result that the final density profile ends up
being very similar to that in the fiducial simulation in which the
halo has a density cusp.

\begin{figure}
	\includegraphics[]{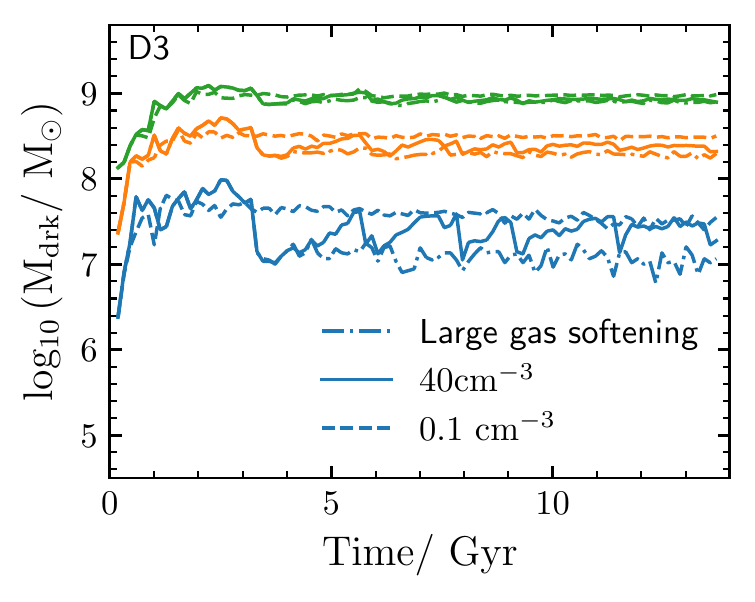}
        \caption{Enclosed dark matter mass within $0.5 \rm \ kpc$,
          $1.5 \rm \ kpc$ and $3 \rm \ kpc$ for simulated dwarf
          D3. From bottom to top different colour lines indicate the
          evolution of the mass within these radii. The dashed lines
          show the result of running the simulation with a density
          threshold of $0.1 \rm \ cm^{-3}$, and the solid lines with a
          threshold of $ 40 \rm \ cm^{-3}$. The dot-dashed lines show
          the result of a similar simulation to that shown by the
          solid lines, but in which the gravity of the gas particles
          is artificially damped by increasing their gravitational
          softening after $t \sim 4 \rm \ Gyr$. This plot demonstrates
          that for dwarf D3 the recovery of the dark matter mass at
          different radii is driven by the collapse of gas.}
    \label{Fig:contration_of_halo2}
\end{figure}

To demonstrate that it is the accretion of baryons following a massive
blowout and subsequent absence of star formation activity that enables
the cusp to recover in the dark matter halo of D3 we carry out another
experiment. Now we stop the simulation at $t\sim 4 \rm \ Gyr$ and
artificially increase the gravitational softening of the gas particles
to a very large value, so that their gravitational force is negligible
compared to that of the dark matter in the inner regions. The result
of this experiment is shown by the dot-dashed line in
Fig.~\ref{Fig:contration_of_halo2}. Clearly, the dark matter halo
remains stable with a shallow density profile when the gravity of the
gaseous component is neglected.

These experiments demonstrate that the main reason for the differences
in the dark matter density profiles of the two dwarfs are the
different timescales on which the potential is perturbed.  The
perturbation timescale is set by the timescale of individual star
formation episodes, which ultimately depends on the timescale on which
gas is able to sink to the centre of the halo. It is then pertinent to
ask: why does it take longer for the gas to collapse and reach the
density threshold for star formation in D3 than in D1?

\begin{figure}
	\includegraphics[]{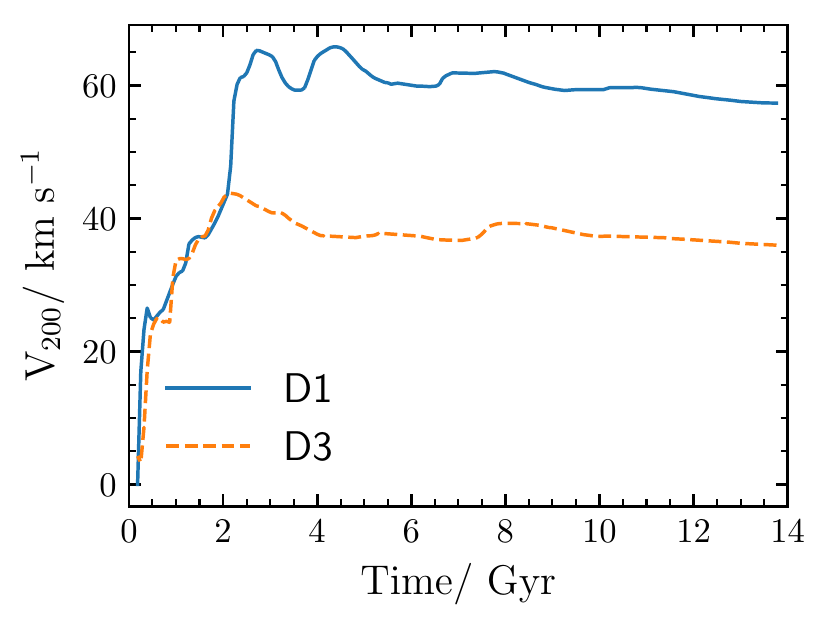}
        \caption{Circular velocity,
          $V_{200} = (G M_{200} / r_{200} )^{1/2}$, as a function of
          time for simulated dwarfs D1 (solid line) and D3 (dashed
          line). At early times both D1 and D3 are identical. After
          $t\sim 2$ Gyr, D1 undergoes a merger that increases its
          circular velocity by more than $50\%$.}
    \label{Fig:circular_velocity}
\end{figure}

The collapse of D1 at early times is essentially indistinguishable
from that of D3. As shown in Fig.~\ref{Fig:origin_cores_all}, their
baryonic (and dark matter) contents before $t \sim 2\rm \ Gyr$ are
comparable at all radii. However, the similarity between the two is
broken soon after. In Fig.~\ref{Fig:circular_velocity} we show the
virial circular velocity, $V_{200} = (G M_{200} / r_{200} )^{1/2}$, of the systems as a function of time. At early times ($t < 2 \rm \ Gyr$),
the virial circular velocities of both dwarfs are almost identical. At
$t \sim 2 \rm \ Gyr$, D1 increases its mass significantly, by a factor
of $3$, and this induces a sudden increase in $V_{200}$ from
$\sim 40 \rm \ km \ s^{-1}$ to $\sim 60 \rm \ km \ s^{-1}$. By
contrast, the virial circular velocity of D3 remains roughly constant
at $V_{200} \sim 40 \rm \ km \ s^{-1}$. After $t \sim 2.5 \rm \ Gyr$
neither D1 nor D3 change their mass or virial velocity
significantly.

The higher mass of D1 at early times enables more baryons to cool and
sink to the centre than in the case of D3\footnote{Note that most of the gas that
  supplies the dwarf galaxies has previously been heated by cosmic
  reionization to a temperature of the order of
  $T \sim 10^{4} - 10^{5} K$, which corresponds to an effective sound
  speed of $c_{s} \sim 10 - 30 \rm \ km/s$. Galaxies whose $V_{200}$
  is comparable to $c_{s}$ are therefore expected to accrete
  systematically less gas than those for which the ratio,
  $V_{200}/c_{s}>>1$.}. This is clear from, e.g.,
Fig.~\ref{Fig:origin_cores_all}, or from the median baryonic profiles
in Fig.~\ref{Fig:density_profiles_D1}. This phase has a strong
impact on the subsequent evolution of the galaxies. The amount
of work that an expanding supernova-heated gas bubble needs to do
against the gaseous halo is much larger in D1 than in D3, simply
because the gaseous halo of D1 is denser. Consequently,
supernova-heated gas travels shorter distances in D1 than in D3. This
is seen in Fig.~\ref{Fig:origin_cores_all}, which shows that baryonic
blowouts push gas out to systematically smaller radii in D1 than in
D3, where the gas is essentially completely ejected from the system
after each episode of star formation. On top of that, the escape
velocity of D1 is larger than that of D3, making it harder for the gas
to escape the system.

The origin of the different degrees of ``burstiness'' in D1 and
D3 is then clear: star formation in D1 proceeds on a much shorter
timescale than in D3 because its gas supply is always much closer
to the centre of the galaxy than in D3, for which gas needs to
travel longer distances before collecting at the
centre of the halo and forming stars. 

\subsection{Discussion}

Our suite of zoom-in simulations demonstrates that baryons can
effectively perturb the inner dark matter density profile of low-mass
galaxies provided the density threshold for star formation is
sufficiently high and supernova feedback is effective at evacuating the
central (gas-dominated) regions. Our numerical experiments indicate,
however, that largely disjoint episodes of star formation, accompanied by
slow reaccretion of gas, leads to dark matter haloes contracting and
erasing the signature of past baryonic blowouts.  

These results help to clarify the emerging view in the literature that
sufficient ``burstiness'' in the star formation activity of dwarfs is
a necessary condition to lower the central dark matter halo density
significantly. Indeed, our simulations indicate that the sudden
removal of gas leads to a substantial reduction in the inner dark
matter content but only when the baryonic component that is removed
had previously been gravitationally significant over an extended
period of time. Burstiness, understood as the rapid assembly and
removal of baryons by star formation, is less effective than a single
large blowout at forming a core, but is necessary to maintain the core
in galaxies that continue forming stars. The inner dark mass may
subsequently recover if baryons gradually reassemble at the center 
over a long period of time. We support these conclusions further,
using a set of idealized experiments, in Appendix~\ref{Sec:App}. 

We therefore conclude that gravitationally-dominant baryonic
perturbations and efficient feedback are both {\it necessary
  conditions} for reducing the inner dark matter density of low-mass
haloes. Experiencing rapid fluctuations in the gravitational potential
is a necessary condition to prevent dark matter haloes from contracting
in response to later accretion of gas. Thus, our simulations
demonstrate that both baryonic blowouts and rapid fluctuations in the
gravitational potential play no role in setting the inner mass content
of dark matter haloes if baryons never dominate the inner gravitational
potential.

Our conclusions have interesting implications. If these mechanism operate in our Universe, some
``core-forming'' dwarfs (like D1) should be currently star-forming and
gas-dominated in the inner regions\footnote{We note that recently,~\cite{DiCintio2017} arrived to a similar conclusion in the context of ultra-diffuse galaxies.}. Some blue compact dwarfs could
fall into this category, as they often exhibit recent or ongoing star
formation and high gas fractions~\citep[e.g.,][]{Tolstoy2009,
  Lelli2014}.

Another implication is that dark matter cores, if they exist at all,
could occur in systems (like D3) that are dark matter-dominated today
and in which star formation shows signs of an abrupt truncation in the
past. Some dwarf spheroidal galaxies could fall into this category if
they were gas-dominated in the past~\citep[see e.g.][]{Weisz2011}.

Finally, we note that simulations by~\cite{Read2016, Read2018} indicate that dark matter cusps may turn into cores even if the baryonic component had not been gravitationally dominant in the inner regions of the halo, in contrast with our results. As shown by~\cite{El-Zant2001}, and more recently by~\cite{Nipoti2015}, the transformation of a central cusp into a core could also be achieved through dynamical friction between dark matter and dense gaseous star-forming clumps that dominate the gravitational potential locally. These gaseous clumps are indeed present in the simulations by~\cite{Read2016}, but not in ours. The differences are likely due to the higher baryon fraction and the lower temperature of the interstellar medium of their simulated dwarfs compared to ours.

\section{EAGLE with higher star formation threshold}
\label{Sec:EAGLE_cores}

The previous section demonstrates that in the EAGLE model of galaxy
formation the inner structure of dark matter haloes is very sensitive
to the assumed gas density threshold for star formation. For our suite
of simulated dwarf galaxies, the size of the core reaches a maximum
for values around $\rho_{\rm th} \sim 50 \rm \ cm^{-3}$ (see
Fig.~\ref{Fig:core_radius}). How do these results affect a larger
sample of galaxies spanning a broader mass range?  We address this
question with a cosmological hydrodynamical simulation performed with
the EAGLE code but assuming this high threshold, which is roughly
$500\times$ higher than the value adopted in the original EAGLE
simulations~\citep[see][]{Schaye2015, Crain2015}.

We have run one dark matter only simulation and two EAGLE
hydrodynamical simulations of a (12~Mpc)$^3$ cosmological volume, one with
the fiducial density threshold for star formation
($\rho_{\rm th} = 0.1 \rm \ cm^{-3}$), and another with
$\rho_{\rm th} \sim 50 \rm \ cm^{-3}$. We shall refer to these
simulations as DMO (dark-matter-only), CLT
(cosmological-low-threshold) and CHT (cosmological-high-threshold),
respectively. The simulated volume is the same as the parent volume
from which our sample of zoom-in dwarfs was drawn, but initialized at
a resolution level comparable to that of our zoom-in suite (resolution
level~L2; $m_{\rm gas} = 6.6 \times 10^4 \ M_{\odot}$).

\begin{figure}
	\includegraphics[]{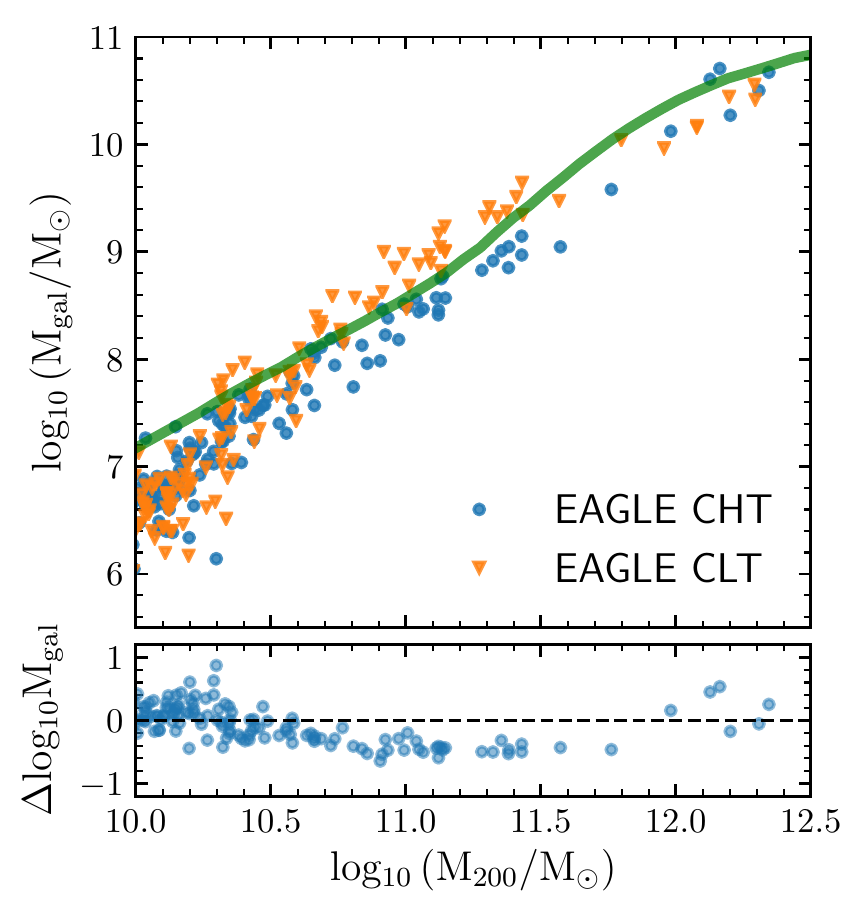}
        \caption{Stellar mass - halo mass relation for galaxies
          identified in the 12~Mpc side-length high-resolution
          (resolution level L2) cosmological hydrodynamical
          simulations with a high density threshold (CHT; blue
          circles) and a low-density threshold (CLT; orange
          triangles). The bottom panel shows the stellar mass ratio
          (CLT to CHT) of the galaxies cross-matched in both
          simulations. The density threshold for star formation
          clearly affects the stellar content of intermediate mass
          haloes ($10^{10.5} < M_{200}/M_{\odot} < 10^{11.5}$), but
          has little impact on the stellar mass of low and high
          mass haloes.}
    \label{Fig:cosmological_cores}
\end{figure}

Galaxies were identified in the CHT simulation as described in
Sec.~\ref{Sec:Simulations} and cross-matched with those identified in
the CLT and DMO simulations, so that {\it all} galaxies have a
counterpart in the other volumes. We do not consider galaxies with
obvious signs of interactions\footnote{We decide whether galaxies are
  undergoing an obvious interaction by visually inspecting them at
  redshift $z=0$. We removed $\sim 30$ galaxies from our sample.},
since these are expected to be out of equilibrium and even the
location of their centres is ambiguous. We include in our sample all
luminous galaxies (i.e., those with at least one stellar particle within the
virial radius) in dark matter haloes of virial mass,
$M_{200} \gtrsim 10^{10} \ M_{\odot}$. This implies that {\it all}
dark matter haloes considered here are resolved with more than
$\sim 25,000$ particles.

The top panel of Fig.~\ref{Fig:cosmological_cores} shows the stellar
mass vs halo mass relation for the galaxies identified in both
hydrodynamical simulations. The green solid lines show the
expectations from abundance matching, as given by
\citet{Behroozi2013}. The bottom panel shows the ratio of stellar
masses in each simulation, as a function of virial mass. Changing the density threshold for star
formation has little impact on the stellar mass of low-mass systems
($M_{200} < 3 \times 10^{10} \rm \ M_{\odot}$) and of high-mass
systems ($M_{200} > 10^{12} \rm \ M_{\odot}$). However, the stellar
content of the intermediate-mass haloes
($5 \times 10^{10} \lesssim M_{200} / \ M_{\odot} \lesssim 10^{12}$)
is significantly affected by the choice of $\rho_{\rm th}$.

In that mass range, CHT galaxies are systematically less massive than
their CLT counterparts (by a factor of $\sim 3$). This systematic
change in stellar mass will have an impact on the galaxy mass
function, which is one of the primary observations used to calibrate
the EAGLE model parameters~\citep{Crain2015}. Since the main purpose
of our exercise is to explore the effects of the threshold on
potential ``cores'' in the inner dark matter profiles, we do not
pursue this isue further here.  However, we note that re-calibration
of the EAGLE model parameters would be necessary to restore agreement
with observations in this regime and there is no gurantee that such
agreement will be achieved.

\begin{figure*}
  \includegraphics[]{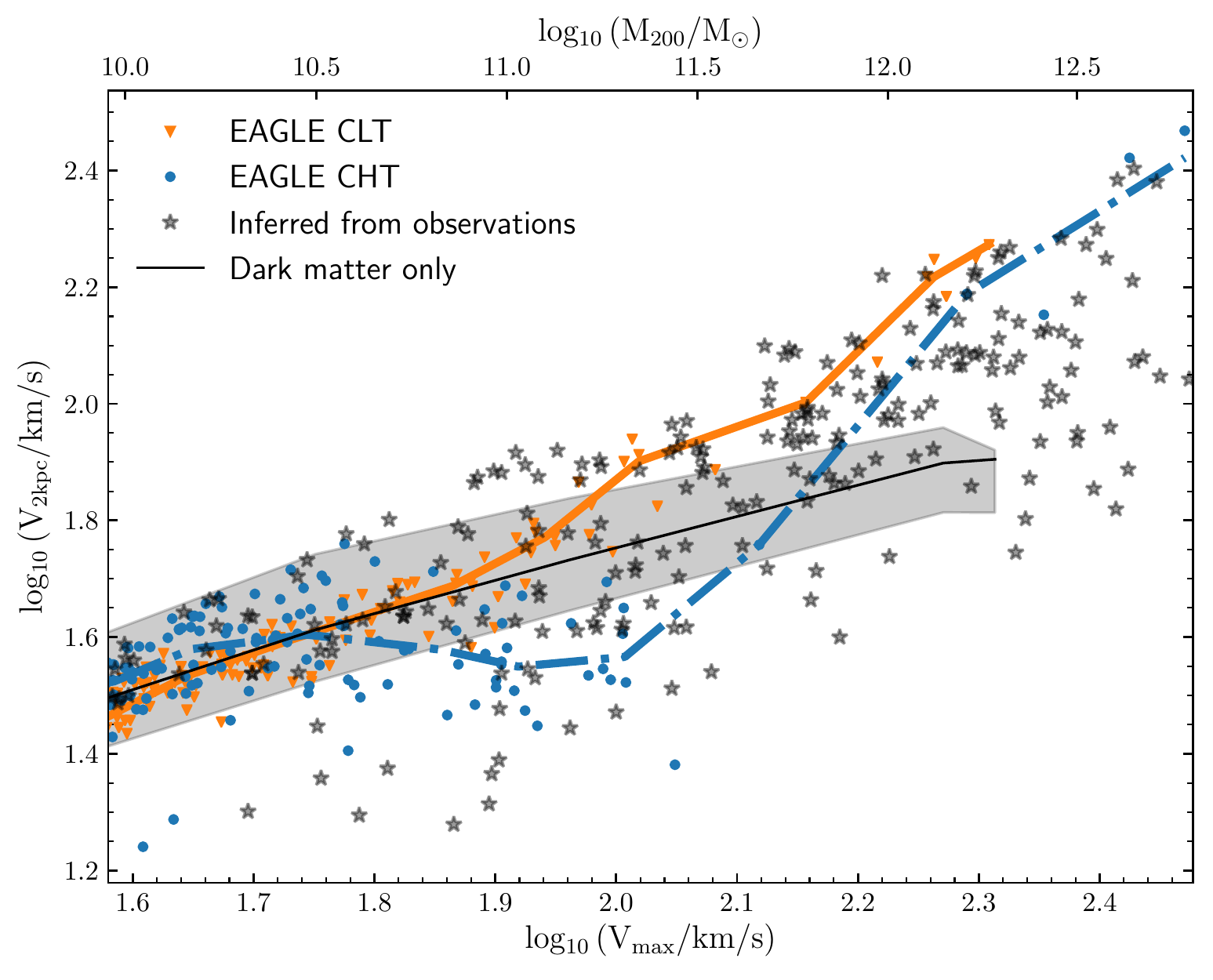}
  \caption{Galaxy circular velocity, measured at $2 \rm \ kpc$ from
    the galaxy centre, as a function of the maximum circular velocity
    (bottom axis) or virial mass (top axis) of the the halo. The
    virial mass shown in the top axis is that of an NFW halo of the
    $V_{\rm max}$ value shown in the bottom axis, and may differ from
    the actual virial mass of the galaxies. Observed galaxies (grey
    starred symbols) were taken from the compilation by
    \citet{Oman2015}. The median and the 10th-90th percentiles of the
    dark matter only simulation are shown by the black solid line, and the
    shaded region, respectively. Galaxies simulated with the EAGLE
    fiducial model (CLT; orange triangles) are roughly consistent with
    NFW haloes at low-mass, but exhibit an excess of mass in the inner
    $2 \rm \ kpc$ at high-mass, owing to the contraction of the halo
    induced by baryons. Increasing the density threshold for star
    formation over the fiducial value has a strong impact on the
    rotation curves. Indeed, for masses
    $M_{200} \gtrsim 3 \times 10^{10} \ M_{\odot}$ (or
    $V_{\rm max} \gtrsim 50 \rm \ km/s$), the simulated galaxy
    population exhibits a significant reduction in $V_{\rm 2kpc}$, in
    contrast with the observational data, which display greater
    diversity. Orange solid and blue dot-dashed lines show the median
    of the CLT and CHT simulations, respectively.}
    \label{Fig:diversity_rotation}
\end{figure*}

\subsection{The diversity of simulated galaxy rotation curves}

A key observable related to the inner mass distribution of galaxies is
their rotation curve. Rotation curves of nearby galaxies display a
great ``diversity'', especially in the regime of dwarf
galaxies. Recently, \cite{Oman2015} have quantified this diversity,
showing that galaxies with similar values of the maximum circular
velocity, $V_{\rm max}$, can have significantly different circular
velocities at $2 \rm \ kpc$ ($V_{\rm 2kpc}$). This is at odds with the
expectation for $\Lambda$CDM haloes, where $V_{\rm max}$ fully
determines $V_{\rm 2kpc}$ (to within a small scatter). 

The origin of this diversity is still not well understood. Part of it may be due to fluctuations in the rotation curves and 
from the fact that $V_{\rm max}$ is approximated, for some dwarfs, by the 
maximum of the observed rotation curve. 
In addition, \citet{Oman2019} argue that some of the diversity may also arise 
from non-circular motions in the gas that are not well accounted for in 
the modelling of the data. Taken at face value, however, 
the data imply a mass deficit in the inner region of some dwarfs compared 
to the expectation from cosmological collisionless N-body simulations.

As shown by~\citet{Oman2015}, the diversity of ``observed'' rotation
curves is at odds with the predictions of the EAGLE and APOSTLE
simulations (which assume a low gas density threshold for star
formation). Indeed, as discussed in the previous section, for low
density thresholds the dark matter distribution in
low-mass systems shows little difference from the dark-matter-only
case. Raising the star formation threshold can in principle reduce the
dark matter central density, but, can it also reproduce the observed
diversity in dwarf galaxy rotation curves?

We explore this in Fig.~\ref{Fig:diversity_rotation}, where we show,
for CLT (in orange, triangles) and CHT (in blue, circles) galaxies,
$V_{\rm 2kpc}$ as a function of $V_{\rm max}$ (or virial mass, scale
on top). The grey band indicates the 10th-90th percentile range
corresponding to the DMO run.  Starred symbols are for
observed galaxies from the compilation by~\citet{Oman2015}. Galaxies
below the grey band are those with less mass within $2$ kpc than
expected from the dark matter alone. This is a quantitative measure of
the reduced inner dark matter density (``inner mass deficit'') that is
associated with ``cores''. Note that galaxies at large masses tend to
have {\it higher} $V_{\rm 2kpc}$ than expected from the dark matter
alone. This is expected, and is just a result of the non-negligible
contribution of the baryons to the inner rotation curve in massive
galaxies.

\begin{figure}
	\includegraphics[]{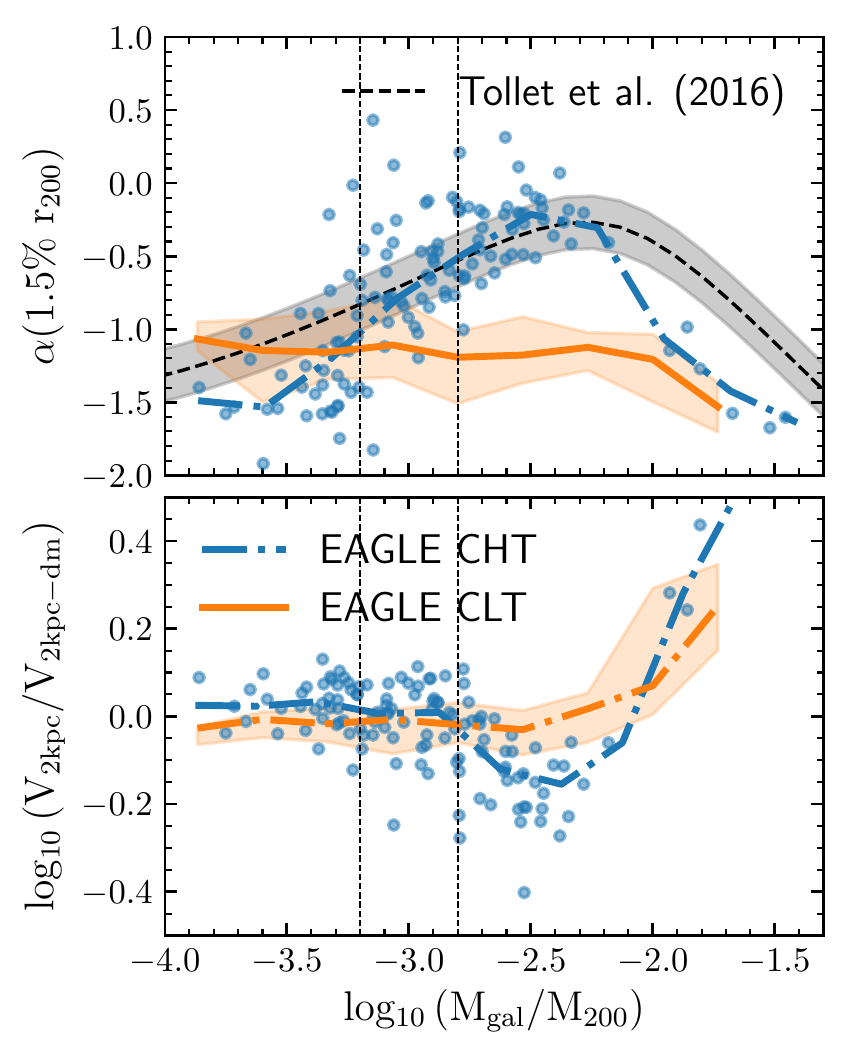}
        \caption{Top panel: the slope of the dark matter density
          profile -- measured at $1.5 \%$ of the virial radius -- as a
          function of the stellar to halo mass ratio,
          $M_{\rm gal}/M_{200}$. Bottom panel: the ratio of the
          circular velocity -- measured at $2 \rm kpc$ -- to the
          median circular velocity at the same radius expected from
          dark matter only simulations (see black solid line in
          Fig.~\ref{Fig:diversity_rotation}), also as a function of
          $M_{\rm gal}/M_{200}$. Individual galaxies from the CHT
          simulation are shown by the blue circles; the median curves
          for the CHT and CLT galaxies are shown by the blue
          dot-dashed and orange solid lines, respectively. The shaded
          orange region encompasses the 10th-90th percentiles of the
          CLT galaxies. The dashed black line (and shaded black region)
          show the fitting formula (and the scatter) that describes
          the inner slope of simulated galaxies from the 
          NIHAO~\citep[e.g.,][]{Tollet2016} and
          FIRE~\citep[e.g.,][]{Bullock2017} simulations.}
    \label{Fig:inner_slope}
\end{figure}

As may be seen from the orange triangles (and the corresponding solid
line tracing the median trend), CLT galaxies follow at low masses the
same trend as the DMO run. In particular, no CLT galaxies 
fall substantially below the grey band, implying that the CLT model
fails to explain the diversity of observed rotation curves and, in
particular, the apparent existence of galaxies with substantial mass
deficits at $2$ kpc.

CHT galaxies, on the other hand, exhibit greater diversity compared to those simulated with the CLT model. They also show pronounced ``mass deficits'' at
$2$ kpc, particularly in the
$60 \lesssim V_{\rm max} / \ {\rm km \ s^{-1}} \lesssim 120$ range,
and seem able to explain some of the observed galaxies with cores in
that range. However, since {\it most} CHT galaxies in the same range
have large deficits, they fail to explain the large number of observed
galaxies that are consistent with a ``normal'' dark matter content in
that range. In other words, a higher EAGLE star formation threshold is
able to modify the inner mass content as a function of $V_{\rm max}$,
but is unable to explain the observed diversity, which demands a
mechanism that creates cores of various sizes in only {\it some}
galaxies, but not in others, over a wide range of $V_{\rm max}$
(roughly between $50$ and $150$ km/s).

\subsection{Comparison with earlier work}
\label{Sec:Tollet_comparison}
The fact that the baryon-induced reduction in central dark matter
density is most effective over a relatively narrow range in mass (as
shown in Fig.~\ref{Fig:diversity_rotation}) has been reported in a
number of recent papers \citep[see; e.g,][and references
therein]{DiCintio2014b,Tollet2016, Bullock2017}. These papers argue
that it is the galaxy formation ``efficiency'' (i.e., the ratio of
stellar-to-virial mass, $\eta_{\rm gal}=M_{\rm gal}/M_{200}$) that
controls the formation of a core. This is, in principle,
reasonable. Galaxies with low $\eta_{\rm gal}$ are unable to form
cores because the few stars that form cannot power the winds needed to
remove baryons from the centre of a galaxy. At the other end, galaxies
with high $\eta_{\rm gal}$ reside in very deep potential wells where,
despite the large amount of feedback energy available, winds are also
unable to remove baryons from the centre.  We now compare some of
those results with ours.

We choose for the comparison the results of \citet{Tollet2016}, who
show that, in their simulations, the slope ($\alpha$) of the dark
matter density profile at $1.5\%$ of the virial radius correlates
strongly with $\eta_{\rm gal}$, reaching a peak of $\alpha\sim 0$ for
$\eta_{\rm gal}\sim 5\times 10^{-3}$ and decreasing for both lower and
higher efficiency values. We compare these results to our CLT and CHT
galaxies in Fig.~\ref{Fig:inner_slope}. As expected, CLT galaxies
show little dependence of $\alpha$ on $\eta_{\rm gal}$, reflecting the
self-similar nature of their largely unperturbed cold dark matter
haloes. CHT galaxies, on the other hand, exhibit a trend in rough
agreement with the results from \citet{Tollet2016}. This is an
interesting result, for it shows that the EAGLE model can, for some
choice of the star formation threshold, reproduce the results of other
simulations where core formation is ubiquitous. We caution, however,
about reading too much into the agreement between our CHT results of
those of \citet{Tollet2016}: the agreement is only crude and
other choices of the threshold would have given different
quantitative results (see Appendix~\ref{Fig:AppAllth_slope}).

A second important point to note is that shallower-than-NFW inner
density profiles at $1.5 \rm \%$ of $r_{200}$ do not necessarily imply low-mass galaxies with a significant mass deficit at $2 \rm \ kpc$, which is what is required to match the observational data shown in Fig.~\ref{Fig:diversity_rotation}. We show
this in the bottom panel of Fig.~\ref{Fig:inner_slope}, where we plot
the change in circular velocity (at $2$ kpc) relative to the
dark-matter-only case. Although most of the CHT galaxies in the range
$-3.2 \lesssim \log_{10} (M_{\rm gal}/M_{200} ) \lesssim -2.8$ (the
range between the vertical dashed lines) exhibit shallower-than-NFW
slopes at $1.5 \rm \%$ of $r_{200}$, they display hardly any mass deficit at $2 \rm kpc$. This is due to the fact that for low-mass systems $2 \rm \ kpc$ becomes an increasingly large distance compared to the size of the dark matter core.

Our conclusion is that baryons {\it can} indeed induce reductions in
the inner dark matter density, but the magnitude of the effect are still unclear, and are heavily dependent on the particular algorithmic choices made in different codes. Although
increasing the star formation threshold in the EAGLE code can result
in cores comparable to those in other simulations, the relatively
narrow mass range of the effect, together with the tight scatter at
fixed mass, result in rotation curves that, taken as an ensemble, do
not compare well with the trends and scatter in existing
data. Modifying the inner cusp over some limited mass range does not
seem to be enough to explain the diversity of dwarf galaxy rotation
curves \citep[see, however,][for a differing view]{Santos-Santos2018}.

\section{Conclusions}
\label{Sec:Conclusions}

We have carried out a set of zoom-in cosmological hydrodynamical
simulations of the formation of low-mass galaxies in the $\Lambda$CDM
cosmology to perform a systematic study of the impact of the gas
density threshold for star formation assumed in the simulations,
$\rho_{\rm th}$, on their inner dark matter content. We have also
simulated a (12~Mpc)$^3$ cubic volume with a high gas density
threshold in order to explore similar baryon effects over a wider
range of galaxy masses. All the simulations of this paper used the fiducial EAGLE model for star formation and feedback, except for varying the gas density threshold for star formation.

Hydrodynamical simulations performed with different numerical codes and/or
different models of galaxy formation often produce different inner
dark matter density profiles.  Although it has long been suspected
that such discrepancies must be related to the way in which subgrid
physical processes are modelled, in this paper we have studied these differences quantitatively. We have tracked down the main differences to details of the modelling of the star formation process, in particular, to the typical density at which gas is turned into stars in a simulation.

If the gas is converted into stars before it has come to dominate the
gravitational potential in the inner regions of dark matter haloes,
then baryonic blowouts do very little to the inner mass distribution
of a dwarf galaxy. By contrast, when the spherically-averaged gas
density becomes comparable to the density of the dark matter halo,
baryon effects can perturb the inner profile. In practice, this can be
achieved by increasing the gas density threshold for star formation to
a value comparable to that of the inner density of dark matter within
the relevant radius. In the EAGLE model density thresholds of order
$\sim 10^{8} M_{\odot} \rm \ kpc^{-3} \sim 10 \rm \ cm^{-3} $ usually
satisfy this condition for a wide range of halo masses (see e.g.,
Fig.~\ref{Fig:density_profiles_D1}). It is then perhaps not surprising
that simulations performed with values of $\rho_{\rm th}$ as high as
this systematically predict shallower dark matter density profiles
than simulations with lower values of this parameter.

Our experiments demonstrate that {\it neither the star formation
  ``burstiness'' nor strong baryonic blowouts are a sufficient
  condition for the inner dark matter content of low-mass galaxies to
  be reduced}. The gaseous content of their haloes must first become
gravitationally dominant in the region of interest, and must be
assembled over a timescale long enough to allow the halo to contract,
before a blowout can flatten the central dark matter density profile.

For dwarfs that are baryon dominated in the inner parts, we
demonstrated that the region of the dark matter halo that is affected
by baryonic blowouts is very sensitive to the value of
$\rho_{\rm th}$. For densities higher than
$\rho_{\rm th} \gtrsim 1 \rm \ cm^{-3}$, we find that the ``core''
radius (or the inner mass deficit) increases roughly monotonically
with $\rho_{\rm th}$ for $M_{200} \gtrsim 3 \times 10^{10} \ M_{\odot}$, but
only for values in the range
$1 \lesssim \rho_{\rm th}/\rm cm^{-3} \lesssim 160$ (see
Fig.~\ref{Fig:core_radius}). For higher densities, supernova feedback,
as implemented in the EAGLE code, is inefficient at driving gas out
from the inner regions. Gas then accumulates at the centre which
becomes compact and overwhelmingly dominated by baryons (see
Fig.~\ref{Fig:Mgal_radius_rho}), causing the dark matter halo to
contract in response. This is
a direct consequence of the reduction in feedback efficiency at high
densities in the EAGLE implementation~\cite[see
e.g.,][]{DallaVecchia2012}. 

The suppresion of gas blowout in baryon-dominated dwarfs is likely to
depend strongly on the specific numerical implementation. For example,
this effect does not seem to be present in simulations of dwarfs
galaxies with the FIRE-2 code~\citep[e.g.,][]{Wetzel2016,
  Fitts2017}. These authors impose very stringent criteria for star
formation on top of a very high value of the gas density threshold for
star formation ($\rho_{\rm th} \sim 1000 \rm \ cm^{-3}$); yet
supernovae explosions seem to be efficient at driving gas out of their
low-mass galaxies.

We have also shown that, for the EAGLE model, global properties of
low-mass galaxies, such as their stellar mass or size are only weakly
affected when increasing the gas density threshold for star formation
from $\rho_{\rm th} = 0.1 \rm \ cm^{-3}$ to
$\rho_{\rm th} = 50 \rm \ cm^{-3}$. However, increasing it beyond
$\sim 50 \rm \ cm^{-3}$ makes supernova feedback less efficient
allowing galaxies to form systematically more stars and become more
compact (see Fig.~\ref{Fig:Mgal_radius_rho}). This implies that
simulations performed with moderate values of $\rho_{\rm th}$ may all
produce similar dwarf galaxy populations even though their inner dark
matter density profile may differ.

For more massive systems ($M_{200} > 5 \times 10^{10} \rm M_{\odot}$),
increasing the density threshold from
$\rho_{\rm th} = 0.1 \rm \ cm^{-3}$ to
$\rho_{\rm th} = 50 \rm \ cm^{-3}$ results in appreciable differences
in galaxy masses. Masses are systematically reduced by a factor of
$\sim 3$ when increasing the value of $\rho_{\rm th}$ from
$\rho_{\rm th} = 0.1 \rm \ cm^{-3}$ to
$\rho_{\rm th} = 50 \rm \ cm^{-3}$ (see
Fig.~\ref{Fig:cosmological_cores}). This is likely to have an impact
on the galaxy stellar mass function which is well reproduced in the
non-core forming EAGLE simulation. Whether
other parameters of the EAGLE model could be adjusted to recover
the good match to the stellar mass function while accommodating cores
in dwarfs remains an open question.

Finally, we have run a cosmological hydrodynamical simulation to
demonstrate that increasing the density threshold for star formation
to $\rho_{\rm th} = 50 \rm \ cm^{-3}$ produces sizeable cores in
essentially {\it all} galaxies in the halo mass range
$5 \times 10^{10} \lesssim M_{200} / M_{\odot} \lesssim 10^{12}$, or
stellar mass range
$10^{8} < M_{\rm gal}/M_{\odot} \lesssim 10^{10}$. This is at odds
with the diverse population of dwarf galaxies observed nearby, where
{\it some} galaxies seem to show cores of various sizes, and over a
wider mass range than in our high-threshold simulation. We emphasize
that these results apply to the EAGLE star formation/feedback
model. It is unclear whether the observed diversity of rotation curves
could be recovered by assuming lower values of the density threshold
for star formation, or by adopting a different numerical prescription
for the feedback processes. 

Our results show that a key parameter that determines whether or not a
core forms in a simulated dwarf is the value of the density threshold
at which gas is turned into stars in the simulation. The size of dark
matter cores also seems very sensitive to this choice, although we find, in addition, that, at fixed value of $\rho_{\rm th}$, the core size at redshift $z=0$ is a function of halo mass (see Fig.~\ref{Fig:density_profiles} and Fig.~\ref{Fig:core_radius}). This is particularly true for intermediate values of $\rho_{\rm th}$ ($1 \le \rho_{\rm th} / \rm cm^{-3} \le 160$), whereas this mass dependence breaks down for lower values of $\rho_{\rm th}$, in which gas is never dense enough to dominate the gravitational potential, and for higher values, for which the EAGLE feedback model becomes inefficient at producing baryonic blowouts. Whether this result holds for other simulation codes deserves further investigation, but if confirmed, it would expose a severe limitation
of current cosmological simulations: they {\it cannot} provide
detailed and reliable quantitative predictions for the inner dark
matter content of low-mass dwarf galaxies. This limitation will be
hard to overcome unless more realistic modelling of the star formation
process is included.

The inner slope of our simulated dark matter haloes -- measured at redshift $z=0$ -- exhibits a mass dependence which is roughly similar to that  reported in earlier work by~\citet[][]{DiCintio2014a} and~\citet[][]{Tollet2016} but with much larger scatter. The larger scatter, although uncertain, might be related to the particular value of $\rho_{\rm th}$ assumed in our work compared to that used by these authors.

If the main mechanisms discussed in our paper do apply in general to real galaxies, then we would expect that 1) ``core-forming'' dwarfs should be
currently gas-dominated and star-forming, as our simulated dwarf D1 is, and 2) dark matter cores
should predominantly occur in systems that are dark matter-dominated
and where star formation shows signs of an abrupt truncation
in the past, as seen in the early stages of evolution of our simulated dwarf D3 (Fig.~\ref{Fig:origin_cores_all}). Some blue compact dwarf and dwarf spheroidal galaxies could be the place where these mechanisms operate. The relation between core formation and galaxy star formation history is, however, less clear; our simulations suggest that, in principle, cusps and cores may exist in galaxies with brief, truncated SFHs, contrary to recent claims~\cite[][]{Read2018}. We caution that these conclusions are drawn from the few specific examples that we have considered in this paper. We are currently carrying out a comprehensive study of the formation of cores in our set of simulated cosmological volumes to see how general these conclusions are and plan to report on this in a forthcoming paper.

\section*{Acknowledgements}
We thank an anonymous referee for useful comments that improved our presentation. We have benefited from the following public {\tt PYTHON} packages:
{\tt NUMPY} ~\citep{Van2011numpy}, {\tt SCIPY}~\citep{Jones2001}, {\tt
  MATPLOTLIB}~\citep{Hunter2007}, {\tt IPYTHON}~\citep{Perez2007} and
{\tt PY-SPHVIEWER}~\citep{Benitez-Llambay2015}. JFN acknowledges the
hospitality of the Aspen Center for Physics, which is supported by
National Science Foundation grant PHY-1607611. This research was
supported in part by the National Science Foundation under Grant
No. NSF PHY17-48958. This work was supported by the Science and
Technology Facilities Council (grant number ST/L00075X/1) and European
Research Council ERC Advanced Investigator grant to CSF, DMIDAS (GA 786910). 
This work used the DiRAC Data Centric system at Durham
University operated by the Institute for Computational Cosmology on
behalf of the STFC DiRAC HPC Facility (www.dirac.ac.uk). This
equipment was funded by BIS National E-infrastructure capital grant
ST/K00042X/1, STFC capital grants ST/H008519/1 and ST/K00087X/1, STFC
DiRAC Operations grant ST/K003267/1 and Durham University. DiRAC is
part of the National E-Infrastructure.




\bibliographystyle{mnras}
\bibliography{biblio} 



\appendix
\section{Short perturbations are less efficient than long perturbations
}
\label{Sec:App}

Fig.~\ref{Fig:origin_cores_all} suggests that long perturbations, in contrast with shorter ones, are more effective at reducing the inner mass of dark matter haloes. We use a set of idealized simulations to demonstrate that this is indeed expected.

We consider a set of numerical experiments in which we perturb an idealized dark matter halo repeatedly with perturbations of different timescales. More particularly, we simulate a perfectly spherical collisionless halo described by a Hernquist density profile~\citep{Hernquist1990}:
\begin{equation}
\rho(r) = \displaystyle\frac{M_{h}}{2 \pi} \displaystyle\frac{a_h/r}{(a_h+r)^3}.
\end{equation}

\begin{figure}
	\includegraphics[]{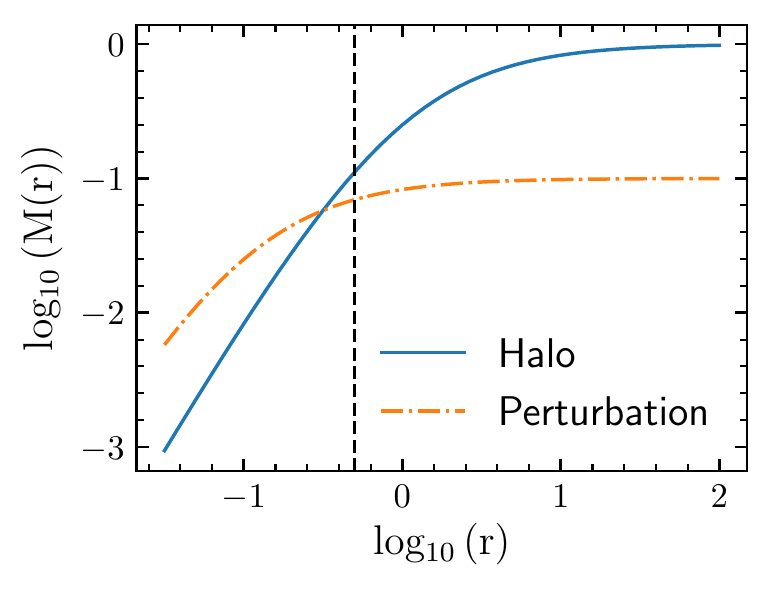}
    \caption{Enclosed mass as a function of radius for the collisionless dark matter halo (blue solid line) and for the external perturbation (orange dashed line). Both, the halo and the external perturbations are modelled by a Herquist density profile. The structural parameters of the halo and the perturbation are $(M_{h}, a_{h}) = (1,1)$ and $(M_{p}, a_{p}) = (0.1,0.1)$, respectively. Vertical dashed line indicates the radius at which we measure the impact of repeated perturbation on the enclosed dark matter mass (see Fig.~\ref{Fig:AppPerturbations}).}
    \label{Fig:AppEnclosedMass}
\end{figure}

The halo, realised with $2\times 10^4$ particles in isotropic orbits, is evolved using the {\tt Gadget-2} code~\citep{Springel2005}. Perturbations are modelled with an external spherical potential also described by a Hernquist density profile. For simplicity we adopt a system of units in which the gravitational constant, the halo scale radius, and the mass are $G=1$, $a_{h}=1$ and $M_{h} = 1$, respectively. The structural parameters of the external perturbations are $a_{p}=0.1$ and $M_{p} = 0.1$. As shown in Fig.~\ref{Fig:AppEnclosedMass}, these spherically symmetric perturbations are gravitationally dominant only in the inner regions of the halo.

\begin{figure}
	\includegraphics[]{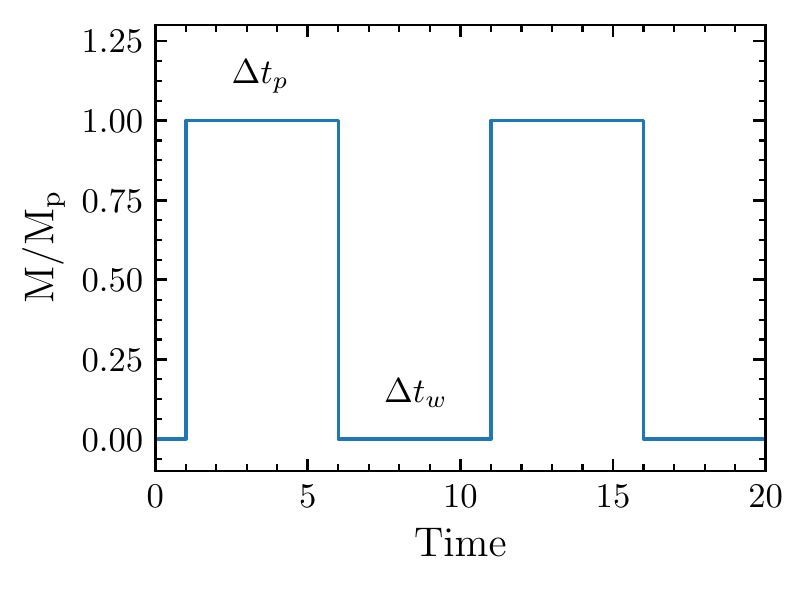}
    \caption{Schematic view of our perturbation scheme. The mass of the perturbation changes instantaneously between $M=0$ and $M=M_{p}$ when the perturbation is turned on and off, respectively. The duration of each perturbation is $\Delta t_{p}$ and the separation between two consecutive perturbations is $\Delta t_{w}$. For our numerical experiments we only consider the particular case in which $\Delta t_{w} = \Delta t_{p}$.}
    \label{Fig:AppSchematic}
\end{figure}

The external perturbations are grown instantaneously, after which they perturb the halo during a timescale $\Delta t_{p}$. After this period of time the perturbation is removed, also instantaneously, and the halo is allowed to relax for a ``waiting'' period of time $\Delta t_{w}$. We only consider the particular case in which $\Delta t_{w} = \Delta t_{p}$. The complete cycle of two perturbations is shown schematically in Fig.~\ref{Fig:AppSchematic}. We note that our perturbation scheme is very particular; results will in general depend not only on the duration of single perturbations, but also on their separation.

The dark matter halo is evolved for a period of time $\Delta t = 150$, which roughly corresponds to the free-fall time of the system at $r=10$. We run 4 simulations varying the frequency of perturbations, which are turned on after $t=4\pi$ (which roughly corresponds to the time required to complete a circular orbit at $r=1$). 

\begin{figure}
	\includegraphics[]{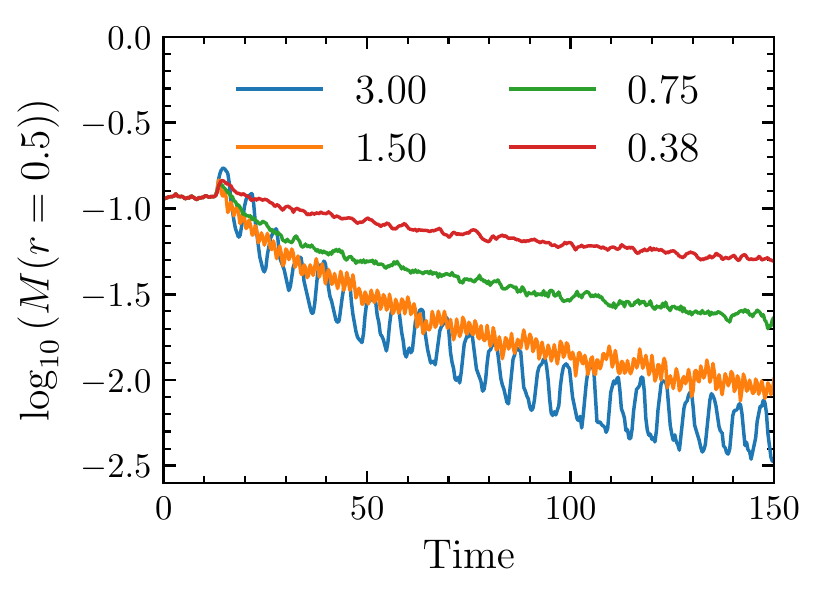}
    \caption{Enclosed dark matter mass within a sphere of radius $r=0.5$, as a function of time. Different curves from top to bottom show the evolution of the enclosed mass while perturbing the system with perturbations of different timescales, as shown in the legend. This experiment demonstrates that having fewer perturbations occurring over longer timescales can reduce the inner dark matter mass of a system much more than having more perturbation operating over shorter timescales.}
    \label{Fig:AppPerturbations}
\end{figure}

The result of these series of experiments is shown in Fig.~\ref{Fig:AppPerturbations}, where we plot the enclosed dark matter mass within a sphere of radius $r=0.5$ -- the free-fall time at this radius is  $t_{ff} \sim 1.5$ -- as a function of time, for different perturbation timescales. Individual perturbations that are active over a long period of time are clearly more effective at driving mass out of the system than shorter ones. Consequently, their integrated effect is also larger. Consider, for example, the perturbations with timescale $\Delta t_{p} = 3$ (blue curve). These are a factor of 2 less frequent than those operating over the longer timescale $\Delta t_{p} = 1.5$ (orange curve), yet their integrated effect, measured at $t=150$, is almost a factor of $3$ larger by comparison. Perturbations acting over timescales much shorter than the free-fall time of the system at this radius, although very numerous, are very inefficient. Note, however, that in {\it all} cases the enclosed dark matter mass decreases secularly and the halo does not contract significantly in response to the perturbations, similarly to what is observed for the dwarf D1 (left panel of Fig.~\ref{Fig:origin_cores_all}). 

\begin{figure}
	\includegraphics[]{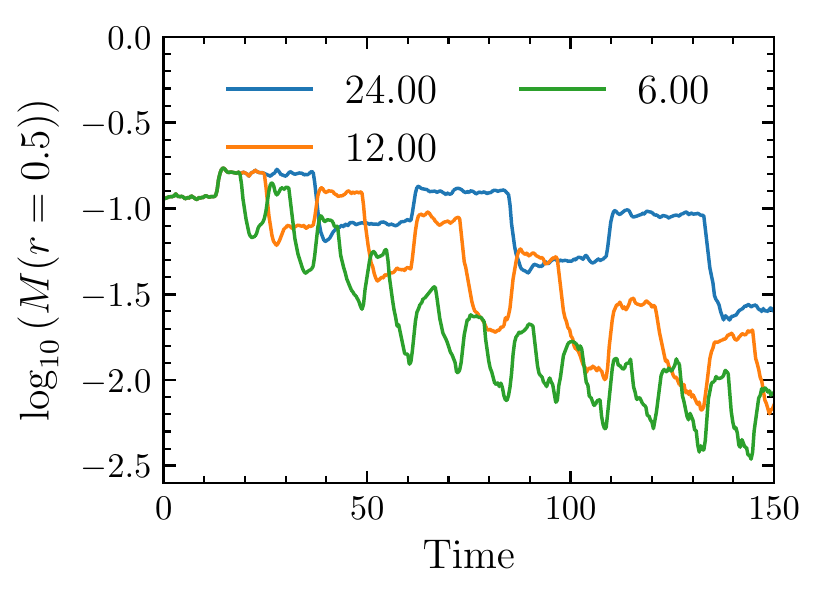}
    \caption{Idem as Fig.~\ref{Fig:AppPerturbations} but for perturbations that are much longer than the free-fall time of the system at $r=0.5$ ($t_{ff} \sim 1.5)$. These series of experiments demonstrate that once the perturbations occur over a long timescale, individual perturbations have a maximal effect to the dark matter halo, but their integrated effect scales monotonically with the number of them.}
    \label{Fig:AppPerturbations2}
\end{figure}

Perturbations that are much longer than the free-fall time are expected to have a maximal individual effect. Therefore, their maximum impact to the enclosed halo mass will ultimately depend on how numerous they are. Fig.~\ref{Fig:AppPerturbations2} shows the cumulative effect of very long perturbations. In contrast to short perturbations, the more numerous the long perturbations are, the greater their integrated impact on the enclosed halo mass. Thus, although longer perturbations are more efficient individually, they must be numerous enough to cause a systematic reduction to the inner mass of the halo. Thus, having very few, separated, and long perturbations, will make the inner dark matter halo to evolve toward states of contraction and expansion, similarly to what is observed for dwarf D3 (right panel of Fig.~\ref{Fig:origin_cores_all}). 
 
We end by noting that the quantitative understanding of these processes requires a comprehensive study of the interplay between the mass distribution of the perturbations, their duration, and separation between them. We are currently performing such a study and will report on this in a forthcoming paper. 

\section{Evolution of the inner dark matter mass for various values of $\rho_{\rm th}$}
\label{Sec:AppB}

\begin{figure*}
	\includegraphics[]{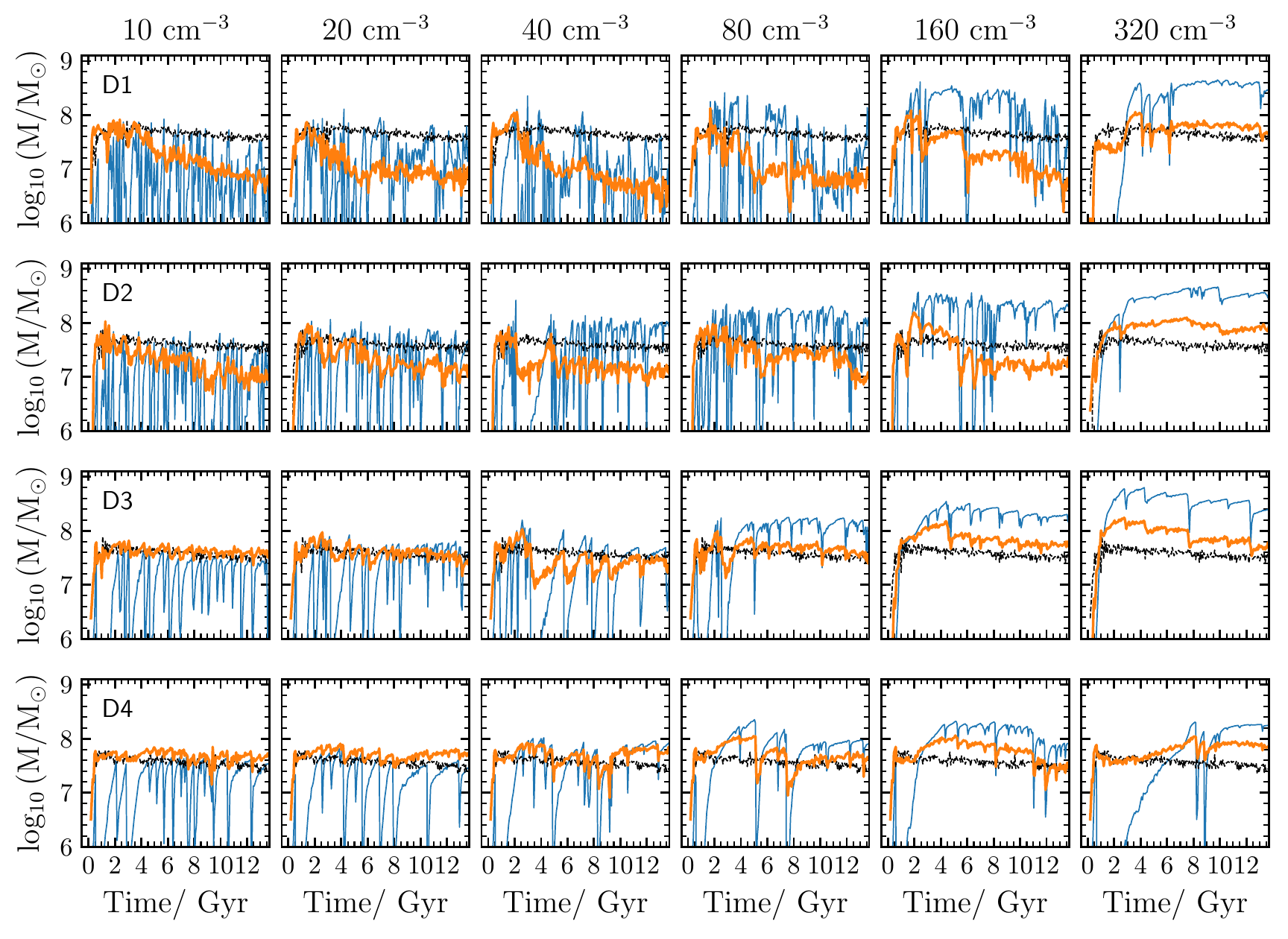}
    \caption{Evolution of the enclosed dark matter mass (thick orange line) and gas mass (thin blue line) at $r=0.5 \rm \ kpc$, for the four zoom-in dwarf galaxies simulated at resolution level L2 (different rows), and for six values of $\rho_{\rm th}$ (different columns). Dashed line shows the enclosed dark matter mass at the same radius, but for the fiducial LT simulation $(\rho_{\rm th} = 0.1 \rm \ cm^{-3})$.}
    \label{Fig:AppAllth}
\end{figure*}

The two examples shown in Fig.~\ref{Fig:origin_cores_all} highlight the main mechanisms of core formation that apply to the EAGLE model. In Fig.~\ref{Fig:AppAllth} we show the evolution of the enclosed dark matter mass within a sphere of radius $r=0.5 \rm \ kpc$ for the four zoom-in dwarf galaxies considered in our work (different rows), and for six different values of $\rho_{\rm th}$ (different columns). The thin blue line shows the enclosed gas mass at the same radius; the black dashed line shows results from the LT fiducial model. 

A number of distinctive trends are seen in this figure. Firstly, as discussed in Sec.~\ref{Sec:Results}, choosing higher values for the density threshold for star formation allows more gas to accumulate in the inner regions of dwarfs before it is turned into stars. This is clearly seen when comparing different columns of Fig.~\ref{Fig:AppAllth}. Note, however, that star formation might occur at much inner radii than $r=0.5 \rm \ kpc$, and therefore the amount of gas mass contained at this radius may not necessarily reflect the actual value of $\rho_{\rm th}$. Secondly, the degree of ``burstiness'' of the dwarfs scales with mass: at a given gas threshold density, the timescale of baryonic perturbations becomes shorter as the dwarfs become more massive. Thirdly, most of the dark matter mass contained at this radius is more efficiently evacuated after a number of distinctive gaseous blowouts. However, dark matter may be recovered on a time interval set by the timescale over which baryons are reaccreted, unless the subsequent accretion and blowouts of gas occurs over a short period of time (``burstiness'').

\section{Relation between the inner slope of the dark matter profile and $M_{\rm gal}/M_{200}$}
\label{Sec:AppC}

In Sec.~\ref{Sec:Tollet_comparison} we compared the inner dark matter profile of our high-resolution cosmological simulations with those of the NIHAO simulations, as reported by~\citet{Tollet2016}. The comparison was done for the specific value of $\rho_{\rm th} = 50 \rm \ cm^{-3}$.  In this section, we perform a similar comparison but for all values of $\rho_{\rm th}$ explored in our work and using all four zoom-in dwarf galaxies.

\begin{figure*}
	\includegraphics[]{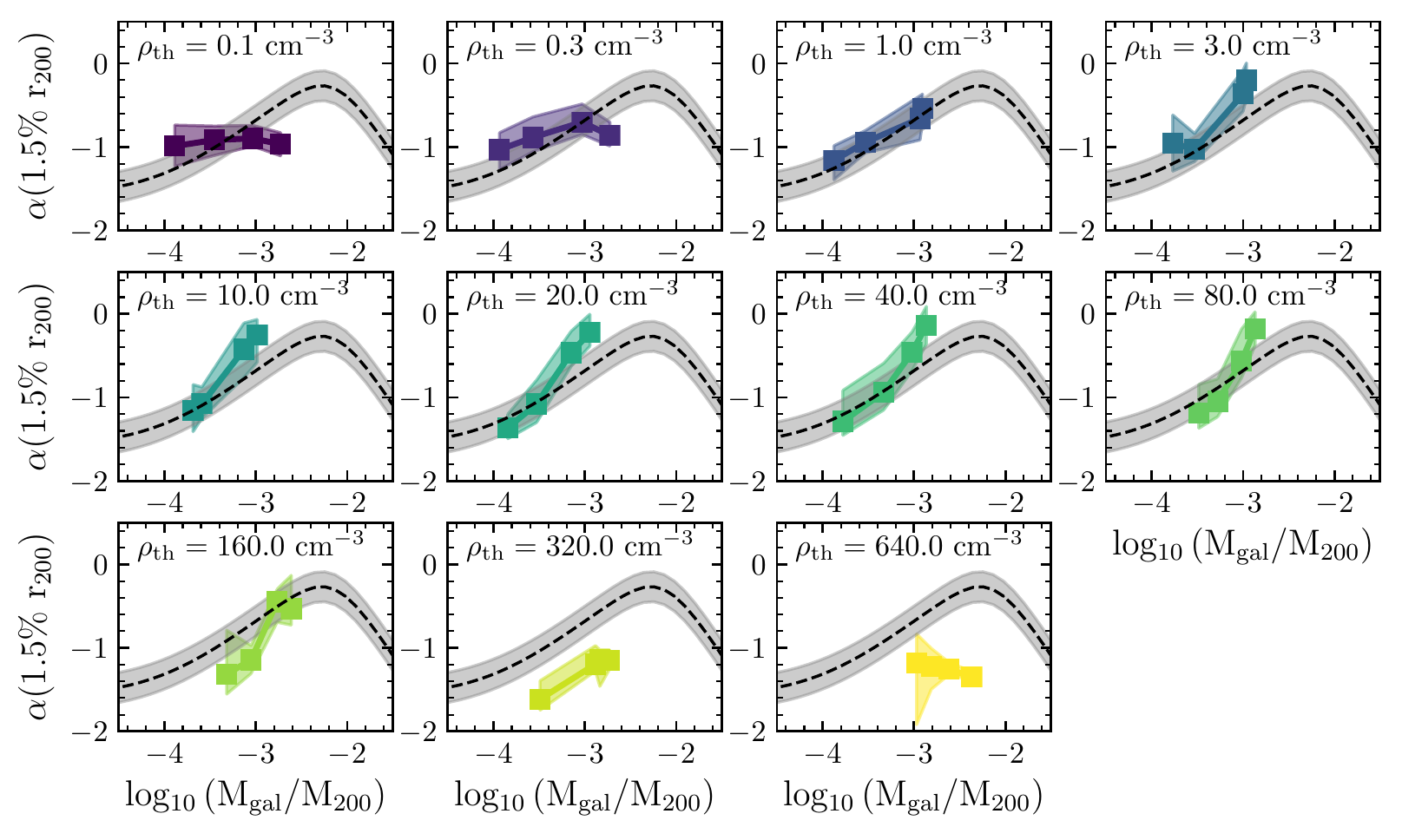}
    \caption{Slope of the dark matter density profile, measured at $1.5 \%$ of the virial radius, as a function of $M_{\rm gal}/M_{200}$, for various values of the density threshold for star formation, $\rho_{\rm th}$. Each symbol represents the median of the four zoom-in galaxies analysed in our work, taken over time using 100 snapshots equally spaced in time, after redshift $z=1$. The shaded regions show the 10th-90th percentiles of the distributions. The dashed line shows the~\citet{Tollet2016} relation, based on the NIHAO simulation suite. The scatter along the relation is shown by the shaded grey band. Our simulations show agreement with NIHAO for relatively low values of $\rho_{\rm th}$. For much higher values, the relation between the inner slope and $M_{\rm gal}/M_{200}$ is steeper than that of NIHAO, except for the two highest values of $\rho_{\rm th}$, for which the dark matter density profiles become very centrally concentrated.}
    \label{Fig:AppAllth_slope}
\end{figure*}

 Fig.~\ref{Fig:AppAllth_slope} shows the inner slope of the dark matter density profile, measured at $1.5 \%$ of $r_{200}$, as a function of $\eta = M_{\rm gal}/M_{200}$, for different values of $\rho_{\rm th}$. Each symbol shows the median for each zoom-in dwarf, taken over time using 100 snapshots equally spaced in time after redshift $z=1$. The coloured shaded regions show the 10th-90th percentiles of the distributions. 
 
 For the lowest values of $\rho_{\rm th}$ (LT simulations), the inner slope is largely consistent with $\alpha = -1$, independently of $\eta$. For $\rho_{\rm th} > 0.3 \rm \ cm^{-3}$, the inner slope of the dark matter profile becomes a function of $\eta$, and agrees well with results reported by~\citet{Tollet2016} (dashed line). This is particularly true for $\rho_{\rm th} = 1.0 \rm \ cm^{-3}$. For values of $\rho_{\rm th} > 1.0 \rm \ cm^{-3}$, the agreement with~\cite{Tollet2016} becomes poorer. In the range $10 < \rho_{\rm th} / {\rm \ cm^{-3}} < 80$, our results also depart from those of NIHAO simulations, although our model exhibits little variation with $\rho_{\rm th}$. For $\rho_{\rm th} > 80 \rm \ cm^{-3}$, the inner slope of all zoom-in dwarfs becomes steeper, as discussed previously. We therefore conclude that, for the EAGLE model of galaxy formation, the relation between the inner slope of the density profile and the $M_{\rm gal}/M_{200}$ ratio is sensitive to the assumed value of $\rm \rho_{\rm th}$.

\bsp	
\label{lastpage}
\end{document}